# A novel linear complementarity approach for analysis of sliding cable with frictions

Ziyun Kan, Haijun Peng*, Biaoshong Chen

*Department of Engineering Mechanics, State Key Laboratory of Structural Analysis for Industrial Equipment, Faculty of Vehicle Engineering and Mechanics, Dalian University of Technology, Dalian, 116024, PR China*

## Abstract

Sliding cable system with frictions is encountered in many engineering applications. Such system is typically characterized by existences of complex and varied motion states of different sliding nodes (pulleys), which leads to significant difficulties for analysis. It is well-known in computational mechanics that complementarity can be advantageously adopted to describe non-smooth phenomenon such as contact-like conditions, quasibrittle fracture, plasticity problems, and these topics have been instigated for many years. Inspired by these existing works, this paper, for the first time, introduces the concept of complementarity in sliding cable modeling. A very simple but effective approach is then developed. Within this approach, the challenging and significant issue of determinations of motion states of sliding nodes, as well as calculations of their sliding lengths, can be handled in a standard linear complementarity formulation, and can be solved using any available efficient solver for linear complementarity problem (LCP). The proposed approach eliminates the need for traditional cumbersome predictor-corrector-based operations. It also opens a way to study the theory problem such as the existence and uniqueness of the solution by using complementarity theories, which is generally absent from existing literature given its complexity. The uniqueness of the solution has been proved in this work, while the existence issue remains open and deserves further investigation. Multi examples involving both static and dynamic analyses are presented to demonstrate the effectiveness of the proposed approach, as well as to reveal some novel phenomenon involved in sliding cable system considering frictions. The results highlight the ability of the method in accurately handling variegated motion states of sliding nodes. The proposed approach, we think, has the potential to be a popular method in dealing with multi-node sliding cable with frictions, given its simplicity and effectiveness.

## 1. Introduction

Even the most complex and advanced structures are built out of certain simple structural components. One of such component is the sliding cable element, which is encountered in many engineering applications, such as domes, cranes, and tensegrity structures. A sliding cable can be considered as one produced by grouping several individual cable elements into a continuous cable. To accommodate this element, some components such as pulleys or sliding contact joints should be installed into the original system. The element can thus, as its name implies, experience relative sliding motions through these structural components. The ability of distributing and transferring load over a long distance through a complex geometric path is probably one of the most significant benefits for using sliding cable elements [1].

Finite element analyses involving sliding cables can be traced back a long time ago and have been performed for various engineering applications. Aufaure [2] pioneered the analysis by developing a three-node finite element for modeling sliding cable passing through a pulley, and subsequently he extended his work to a clamp-cable element [3]. Zhou et al. [4] developed a similar three-node finite element using the total Lagrange formulation and employed such element into analyses of airdrop systems. Lee et al [5] proposed a finite element model for modelling specific hinges such as doubled sling and butterfly loop of sling systems involving cable sliding effects. These early works all focus on the three-node model, assuming that only one pulley is composed in the element. To analyze a sliding cable containing many pulleys, one can tentatively use these models, which is done by adding some assistant nodes into each interior segment such that a continuous cable can be resolved into a certain number of three-node elements, as indicated in Fig. 1. However, such a treatment has two inherent drawbacks: on one hand, it enlarges the solving dimension due to these additional assistant nodes, and on the other hand, cumbersome remeshing operations [6] must be adopted if any assistant node has passed through a pulley.

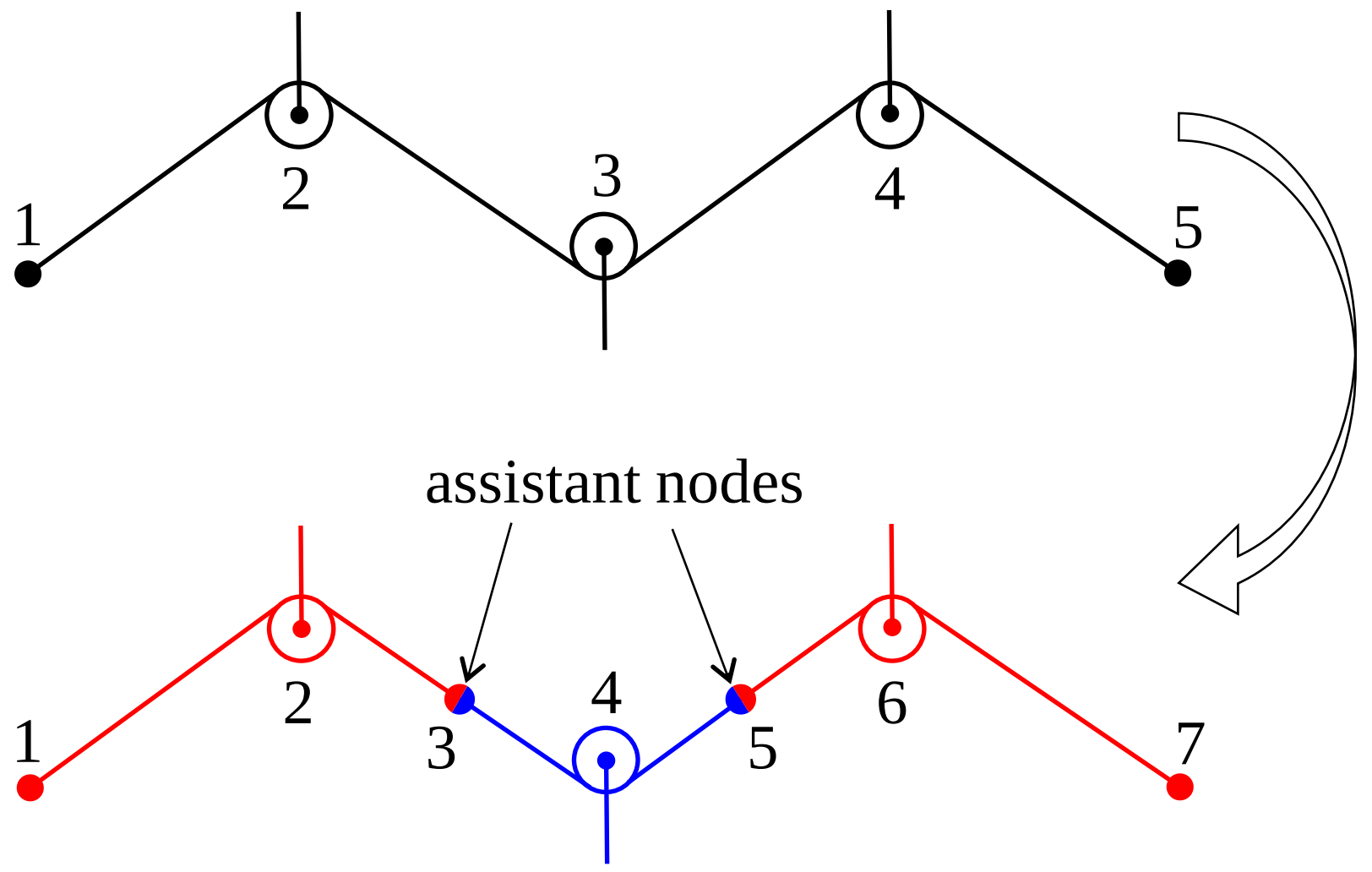

Fig. 1 Transforming a five-node sliding cable into three three-node elements by introducing two assistant nodes.

A more straightforward way dealing with sliding cables containing many pulleys is to directly formulate a multi-node element. Ju and Choo [1] proposed a super element for small displacement static analysis of cable passing through multiple pulleys. Chen et al. [7] developed a multi-node sliding cable element for static analysis of suspen-dome structures. A recent interest also emerges in actively control of tensegrity structures using the concept of sliding cables. To analyze such kind of structure, sliding cables must be special handled. Moored and Bart-Smith [8] first proposed the terminology “clustered tensegrity” to denote those tensegrity structures actuated by sliding cables (clustered actuation strategy). They also systematically studied the prestress and stability performance of the structure based on an energy approach. Bel Hadj Ali et al. [9] presented a modified dynamic relaxation approach to take into account cable sliding effects in analyzing the deployment performance of clustered tensegrities. Zhang et al. [10, 11] employed a co-rotational formulation for modeling sliding cables in geometric nonlinear analysis of clustered tensegrities. Previous studies of clustered tensegrities are all conducted from a static point of view. Recently, Kan et al. [12, 13] investigated the dynamic deployment performances of clustered tensegrities using different modeling methods.

All the above mentioned works, except the ones in Lee et al [5] and Ju and Choo [1], involving sliding cable analyses adopt a frictionless assumption for modeling cable passing through pulleys or other sliding joints. The frictionless model assumes an uniform tension for different segments of the cable, and the tension can be simply computed from the whole deformation of the cable in combination with its constitutive relation [7, 13], without any special consideration of, such as, whether a pulley is sliding or not. This assumption is able to capture the general sliding behavior between a cable and the involved pulleys, and it largely simplifies the formulation in analyses. However, the frictionless assumption is merely an idealized concept physically, and may results in considerable deviations from reality. This has been demonstrated by experiments in available literature. Liu et al. [14] investigated a pre-stressing of suspen-dome structure equipped with sliding cables, and the outcomes showed that the tension deviations from values determined using the frictionless model exceeded 30% for some cables. Veuve et al. [15, 16] studied the deployment performance of a tensegrity footbridge having sliding cables. Results demonstrated that frictions have a significant impact on the motion of such structure. Similar conclusions were also obtained in other engineering applications involving sliding cables [1, 17]. It appears that the consideration of frictions is critical for modeling system behaviors accurately, although it will largely increase the complexity of analysis. To address frictions, many efforts have been done over the past years [1, 5, 17-19]. Very recently, Bel Hadj Ali [20] extended the original dynamic relaxation method to analyze cable structures taking into account the sliding-induced frictions. Coulibaly [6] developed an explicit dynamic formation sliding cable element from a comprehensive view of point. In contrast to frictionless case, most formulations involving frictions consider non-uniform tensions in different segments, and tensions in every two adjacent segments should in agreement with the permitted range defined by sliding criterion.

One of the extremely important issues in the analysis, which seems have not be fully emphasized in the available literature, is the determination of motion states of the pulleys, because it is closely related to calculations of sliding lengths of each pulley. This is essentially equivalent to answer the following questions. For every single pulley, is such pulley sticking? If not, which side does it slide towards? Most formulations [1, 6, 17, 18] using a predictor-corrector-based approach to deal with this issue. In the analysis, when the estimated nodal coordinates or displacements are given in an iteration step, the predictor-corrector-based approach assumes that all the pulleys have been stuck for the first prediction, then the tension in each segment can be calculated based on constitutive relations. If tensions in two sides of a pulley exceed the permitted range defined by sliding criterion, then the pulley is assumed having slid, and the tensions should be adjusted to meet the *limit state* of the sliding functions in the next correction. This process repeats until all the tensions are in the permitted range. Generally, this process works well for a sliding cable element composed of only a few pulleys because the total ergodic states are acceptable, and it is reasonable to think that solutions can be found in only a few correction operations. This is however not the case for a sliding cable with a great number of pulleys. Considering a sliding cable having $N$ pulleys, as each pulley has three possible motion states (slide towards either side or stick state), there are $3^N$ possible states in total, the number quickly get unmanageable as $N$ increases. In [6], the model is formulated by first dividing a continuous multi-nodes sliding cable into a number of groups, and each group is composed of these nodes that share the same slide/stick states, then a one-step correction is performed independently on every single group (of slide state) using Newton–Raphson method. Such a treatment splits the relationship between different parts of the organic sliding cable element, and the results may not strictly guarantee that the intersect node of two adjacent groups satisfying the sliding criterion as well. In general, the predictor-corrector-based approach used in existing literature is not only cumbersome but also inefficient as the convergence cannot be ensured. It cannot generally meet the enormous computing requirements for time history dynamic analysis, typically characterized by thousands of time steps, where repeated analysis is needed for every single sliding cable in every single time step (or even iteration step, if an implicit algorithm is used). Besides, most formulations involving sliding cable with frictions focus purely on a numerical aspect. Theoretical analyses such as the existence and uniqueness of solutions of sliding lengths of the pulleys are rarely seen. Inquiries to these theoretical issues may provide us a profound insight into the sliding behaviors of the system.

This paper directly addresses the above mentioned drawbacks of traditional analysis method. The objective is to develop a linear complementarity approach for analysis of multi-node sliding cables with frictions. It is well-known that complementarity has been extensively investigated in a large variety of economic and engineering applications [21]. In particular in structural mechanics, complementarity can be advantageously adopted to describe contact-like conditions[22-24], quasibrittle fracture [25], plasticity [26, 27] and the unilateral behavior of cables [28, 29]. The present work is exactly inspired by these existing works. Within the proposed complementarity approach, the previously mentioned challenging problem of determination of the motion states of pulleys, as well as calculations of the involved sliding lengths, can be handled in a very simple standard linear complementarity formulation. The solutions thus can be obtained by using mature algorithms having been developed in the research field of computational mathematics. These algorithms are typically available as black boxes for general engineering applications. In contrast to cumbersome traditional methods, the proposed approach allows us to write the program in a very compact form. In addition to these merits from the numerical aspect, the proposed approach also paves a way for investigations of theory problems such as the existence and uniqueness of the solution by using complementarity theories, which are typically absent from existing literature given their complexity, as mentioned before. Such problems are partially investigated in this work, yet have not been fully completed. The rest of this paper is organized as follows. Some brief reviews of linear complementarity theory relevant to the current work are first presented in Section 2. Section 3 gives a detailed description of the presented linear complementarity approach in analysis of a sliding cable with frictions. Multi examples involving both static and dynamic analyses are presented in Section 4 to demonstrate the effectiveness of the proposed approach. Finally, some conclusions are summarized in Section 5.

## 2. Brief review of linear complementarity theory

Before presenting the proposed approach for sliding cable analysis, some brief reviews of linear complementarity theory is given in this section.

### 2.1 Definitions of the linear complementarity problem

A linear complementarity problem (LCP) is the problem of finding the solutions of $\boldsymbol{z} \in \Re^n$ subjected to the following equality and inequality conditions:

$$\begin{cases} \boldsymbol{w} = \boldsymbol{M}\boldsymbol{z} + \boldsymbol{q} \\ \boldsymbol{w} \geq \boldsymbol{0}, \boldsymbol{z} \geq \boldsymbol{0}, \boldsymbol{w}^{\mathrm{T}}\boldsymbol{z} = 0 \end{cases} \tag{1}$$

where $\boldsymbol{q} \in \Re^n$ is an $n$-dimensional constant column vector and $\boldsymbol{M} \in \Re^{n\times n}$ is a given square matrix of dimension $n$, and the terms $\boldsymbol{w} \geq \boldsymbol{0}$ and $\boldsymbol{z} \geq \boldsymbol{0}$ mean that each component of $\boldsymbol{w}$ and $\boldsymbol{z}$ is nonnegative. The above LCP is denoted by the pair $(\boldsymbol{M}, \boldsymbol{q})$.

The inequality complementary condition in expression (1) is commonly written in a compact form:

$$\boldsymbol{0} \leq \boldsymbol{w} \perp \boldsymbol{z} \geq \boldsymbol{0} \tag{2}$$

where $\boldsymbol{w} \perp \boldsymbol{z}$ denotes $\boldsymbol{w}^{\mathrm{T}}\boldsymbol{z} = 0$. According to the nonnegative characteristic of $\boldsymbol{z}$ and $\boldsymbol{w}$, the complementary relationship $w_i z_i = 0$ holds for every single component. An LCP can have a unique solution, multiple solutions or no solution at all. From the above complementary relationship, it follows that when $w_i > 0$, then $z_i = 0$, and vice versa. This implies that one can use an exhaustive method to find out all existing solutions, which treats the problem by a combinatorial evolution of the complementary relationship $w_i z_i = 0$. An LCP of dimension $n$ provides $2^n$ different combination cases.

For a problem of large dimension, the exhaustive method becomes numerically expensive as $2^n$ grows rapidly. A more efficient approach is using the complementary pivot algorithm, commonly referred to as Lemke's algorithm [30]. Other efficient algorithms to solve LCP can be found in [21]. If special LCP algorithms are not available, one can also use a optimization algorithm to find the solution, which can be done by transforming the original $\mathrm{LCP}(\boldsymbol{M},\boldsymbol{q})$ into the following quadratic programming:

$$\begin{aligned} &\text{minimize} \quad \boldsymbol{z}^{\mathrm{T}}\left(\boldsymbol{Mz}+\boldsymbol{q}\right) \\ &\text{subject to} \quad \boldsymbol{Mz}+\boldsymbol{q}\geq \boldsymbol{0}, \boldsymbol{z}\geq \boldsymbol{0} \end{aligned} \tag{3}$$

If $\mathrm{LCP}(\boldsymbol{M},\boldsymbol{q})$ have a solution, such solution must be the solution of the quadratic programming (3) as well, and it corresponds to a target value zero.

2.2 The ***w***-unique property of an LCP

In addition to the numerical aspect, theoretical analysis concerning with the existence and multiplicity of solutions is another topic in research field of LCP. Over the past five decades, fruitful achievements have been made on this topic. Here, we only touch on some basic definitions and conclusions relevant to the present study. Interested readers can refer to monograph [21] for deeper and more comprehensive theories.

**Definition 1:** The LCP is said to be *solvable* if it has at least one solution. The $\mathrm{LCP}(\boldsymbol{M},\boldsymbol{q})$ is said to be ***w****-unique* if any two solutions $\boldsymbol{z}^1$ and $\boldsymbol{z}^2$ give rise to the same vector $\boldsymbol{w}=\boldsymbol{Mz}^i+\boldsymbol{q}$ ($i=1,2$).

**Definition 2:** A matrix $\boldsymbol{M}$ is said to be a $\boldsymbol{P}_0$-matrix if all its principal minors are nonnegative. The class of such matrices is denoted by $\boldsymbol{P}_0$.

The following theorem identifies a class of matrices $\boldsymbol{M}$ for which all solutions of a *solvable* $\mathrm{LCP}(\boldsymbol{M},\boldsymbol{q})$ must be ***w****-unique*. Elegant proof of such theorem can be found in [21].

**Theorem:** if $\boldsymbol{M}$ is a $\boldsymbol{P}_0$-matrix and for each index set $\alpha$ with $\det \boldsymbol{M}_{\alpha\alpha}=0$, the columns of $\boldsymbol{M}\left(:,\alpha\right)$ are linearly dependent, then $\mathrm{LCP}(\boldsymbol{M},\boldsymbol{q})$ is ***w****-unique*.

## 3. Formulation of sliding cable with frictions

The formulation of a sliding cable element with frictions is presented in this section, in combination with the linear complementarity approach. First, it is important to clarify some modeling assumptions used in this work, as follows:

(1) Pulleys are small enough compared with the feature scale of the sliding cable. A pulley cannot slide across the adjacent pulley along the cable, which means that orders of pulleys along the cable remains unchanged.
(2) The sliding cable is perfectly flexible such that bending effects can be neglected and every segment of the cable remains straight.
(3) Deformation of the cable is considered small. Linear elastic constitutive relation is considered. Stress and strain are constant over every segment.

It is worth to emphasize that the small deformation assumption in (3) only means that the axial elongation of the cable (as well as that of each segment) is small compared to its original length; it does not necessarily require that the whole structure is subjected to a small displacement, and sliding length of each pulley can also be arbitrarily large (in the admissible range). Based on the above assumptions, the sliding cable in a structure can be defined as a set of continuous segments which can be fully described by position vectors of the involved pulleys. Without loss of generality, the proposed formulation is first presented in the context of an explicit dynamic analysis, based on position descriptions. Extensions to implicit dynamic and static analysis will be discussed subsequently.

For a structure containing sliding cables subjected to external loads, the governing equation of dynamic analysis is mathematically formulated as a set of differential equations. Time discretization strategy is necessary to obtain its numerical solution. Let us consider the analysis of current time $t$ using a time step size $\Delta t$. At the previous time $t-\Delta t$, every information is known. These information includes but not limited to basic variables of the governing equation (i.e., position, velocity and acceleration of every node). For explicit dynamic analysis at the current time step, the position vector of every node can be first updated using the velocity and acceleration information of the *previous step* with the time step size $\Delta t$. The rest core task is exactly to determine the nodal force vectors under the *newly updated* nodal position vectors (configuration). Once these nodal force vectors are obtained, the nodal velocity and acceleration vectors can be updated using any explicit algorithm. This completes analysis of the current time step.

### 3.1 General formulation of a sliding cable element

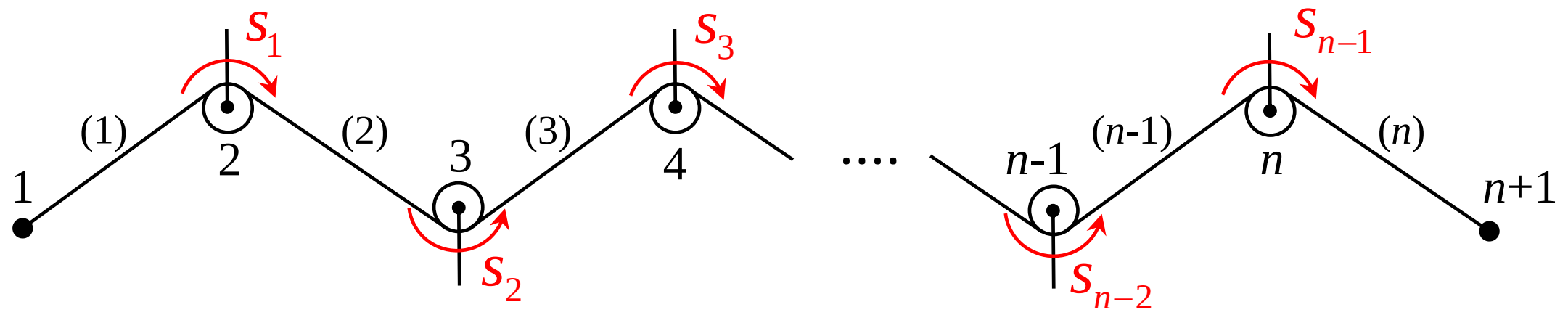

Fig. 2 A multi-node sliding cable element.

Let us directly focus on the above mentioned core task in analysis of a single sliding cable. Consider the sliding cable element shown in Fig. 2. The element is composed of $n$+1 nodes, $n$ segments and $n$-1 frictional pulleys. The nodal indexes are arranged in order for conciseness. At the current analysis step of time $t$, the nodal positions are collected as

$$\boldsymbol{q}_{\mathrm{s}} = \left[\boldsymbol{R}_1^{\mathrm{T}}, \boldsymbol{R}_2^{\mathrm{T}}, \ldots, \boldsymbol{R}_n^{\mathrm{T}}, \boldsymbol{R}_{n+1}^{\mathrm{T}}\right]^{\mathrm{T}} \in \Re^{3(n+1)} \tag{4}$$

where $\boldsymbol{R}_i = [x_i, y_i, z_i]^{\mathrm{T}}$ represents the position vector of the $i$th node. $\boldsymbol{q}_{\mathrm{s}}$ can be interpreted as the generalized coordinates which describes the configuration of this element. Define $\boldsymbol{l}_i$ , $l_i$ and $\hat{\boldsymbol{l}}_i$ , respectively, as the current vector, vector length and vector direction corresponding to the $i$th segment ( $\overrightarrow{N_{i+1}N_i}$ ($1 \le i \le n$)) of the element. It is easy to compute that

$$\boldsymbol{l}_i = \boldsymbol{R}_i - \boldsymbol{R}_{i+1} = [x_i - x_{i+1}, y_i - y_{i+1}, z_i - z_{i+1}]^{\mathrm{T}} \tag{5}$$

$$l_i = \sqrt{\boldsymbol{l}_i^{\mathrm{T}} \boldsymbol{l}_i} = \sqrt{(x_i - x_{i+1})^2 + (y_i - y_{i+1})^2 + (z_i - z_{i+1})^2} \tag{6}$$

$$\hat{\boldsymbol{l}}_i = \boldsymbol{l}_i / l_i \tag{7}$$

The formulation of $\hat{\boldsymbol{l}}_i$ given above denotes the tension direction of the $i$th segment at the current time. The rest task is exactly to obtain the tension size of the segment. For frictionless case, as tensions in all segments are exactly the same, one can directly sum the current length of every segment and then minus the free/non-loaded length of the whole element to obtain its axial deformation. Tensions of the whole element (as well as that of each segment) can be computed straightforwardly using its constitutive relation. The details, such as whether a pulley has been slid or not, or which side (direction) does it slide towards if the sliding motion indeed occurs, are unnecessary to be concerned with. From another aspect, calculations in frictionless case are independent of the loading path; that is to say, the nodal force vectors can be fully determined by the current state of the nodal positions, without any historical information of the past time steps. This is, however, not the case for analysis when frictions are considered, in which tensions of different segments may be no longer the same, and one cannot simply obtain the tensions according to the total deformation, as it does in the frictionless case. Rather than treating the sliding cable as an integral element, it is more likely that in this case each segment can be considered as an individual classical cable element. The analysis is dependent of the loading path and one must carefully track the sliding behavior of each single pulley in each time step, as detailed below.

Considering the analysis from the previous time step to the current time step, the whole cable element may experience simultaneously the sliding motions of some pulleys and the deformations of all segments. Such process is hybrid. To handle it, an important consideration in the present work is treating it successively: we first deal with the sliding motions of the pulleys, then the deformation of every segment can be obtained by considering the pulleys are fixed to the cables. In the first process, the sliding motions of pulleys directly govern the *non-loaded* length of every segment.

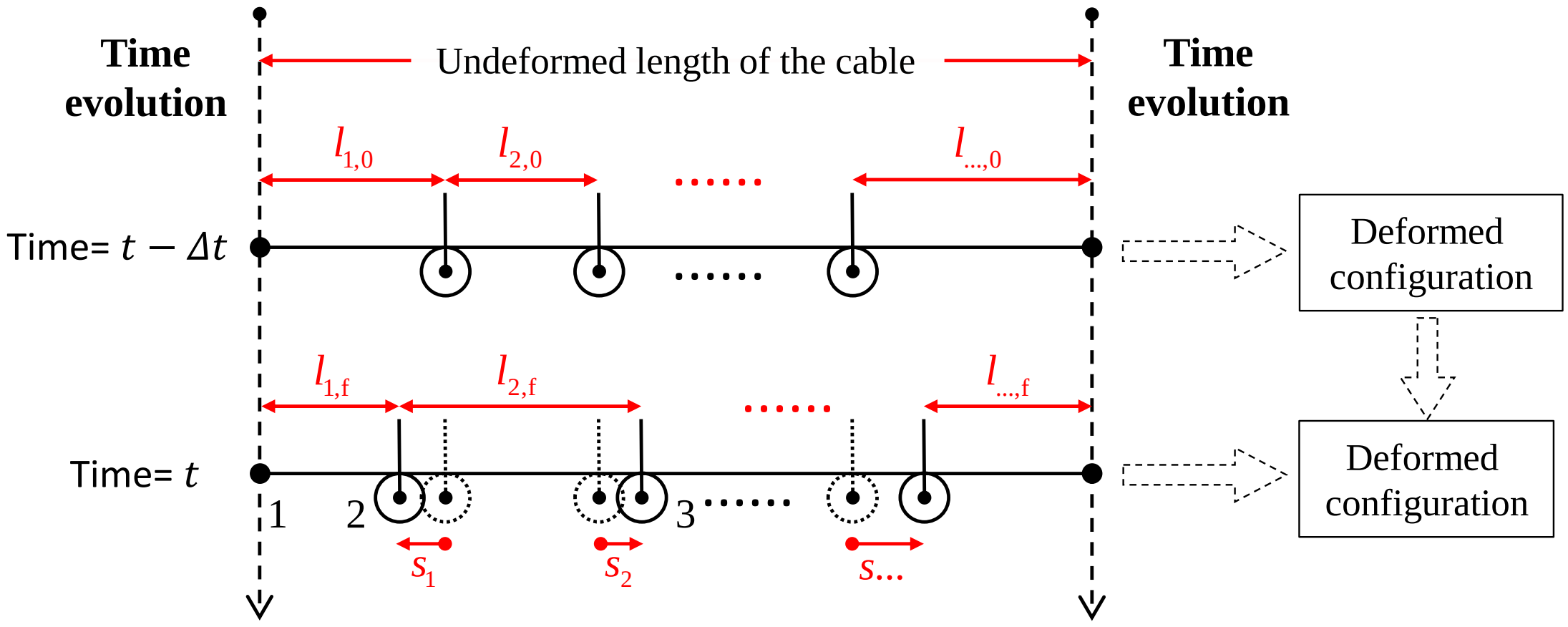

Fig. 3 Sliding motions of pulleys with time evolution.

Since every information is known at the previous time step, let us denote $l_{i,0}$ as the *non-loaded length* of the *i*th ($1 \le i \le n$) segment at the previous time step. Introduce an *unknown* sliding variable $s_i$ ($1 \le i \le n-1$) for each pulley, where $s_i$ represents the sliding distance of the *i*th pulley towards the end (*n*+1th) node from the previous time step to the current one.A negative value of $s_i$ represents the pulley has slid towards the start (first) node, and apparently, $s_i = 0$ represents no sliding motion occurs and the pulley is sticking (the material point is clinging to the cable). As indicated in Fig. 3, all the sliding distances are measured with respect to the *undeformed* configuration. The *non-loaded* length of each segment at the current time step thus can be updated by taking into account the sliding motions.

$$l_{i,\mathrm{f}} = \begin{cases} l_{1,0} + s_1 & i = 1 \\ l_{i,0} - s_{i-1} + s_i & 2 \le i \le n-1 \\ l_{n,0} - s_{n-1} & i = n \end{cases} \tag{8}$$

where $l_{i,\mathrm{f}}$ denotes the *non-loaded length* of the *i*th segment at the current time step. The axial elongation of the *i*th segment yields:

$$\Delta l_i = l_i - l_{i,\mathrm{f}} \tag{9}$$

According to the small deformation assumption ($\Delta l_i \ll l_i \left( l_{i,\mathrm{f}} \right)$), the axial stiffness of the *i*th segment can be approximated as:

$$k_i = \frac{EA}{l_{i,\mathrm{f}}} \approx \frac{EA}{l_i} \tag{10}$$

where *E* and *A* are the elastic modulus and the cross-sectional area of the cable. The tension of the *i*th segment can be simply written as:

$$T_i = k_i \Delta l_i \tag{11}$$

By taking into account the tension directions, the nodal force vector applied to the *i*th node eventually yields:

$$\boldsymbol{f}_i = \begin{cases} -T_1\hat{\boldsymbol{l}}_1 & i=1 \\ T_{i-1}\hat{\boldsymbol{l}}_{i-1} - T_i\hat{\boldsymbol{l}}_i & 2 \le i \le n-1 \\ T_n\hat{\boldsymbol{l}}_n & i=n+1 \end{cases} \tag{12}$$

So far, the only unknowns in Eq. (12) are the sliding distances of the pulleys, and the rest task is exactly to find the solutions of them.

### 3.2 Calculations of sliding distances using linear complementarity approach

To determine the sliding distance of every pulley, sliding criterion must be introduced. As adopted in many existing literature [1, 5, 6, 20], the well-known capstan equation is used as the sliding criterion, which states the following relationship:

$$T_2 = \mathrm{e}^{\mu\theta} T_1 \tag{13}$$

where $T_1$ and $T_2$ are the tensions of two sides of a pulley, and $\theta$ and $\mu(\ge 0)$ are, respectively, the total contact angle (as indicated in Fig. 4) and the friction coefficient between the cable and the pulley. Eq. (13) defines (for the case of $T_2 \ge T_1 > 0$) the sliding limit state (the verge of sliding or sliding) of the tensions of two sides. Detailed derivations of the equation can be found in [31]. One should keep in mind that the equal sign in expression (13) may not always hold: it only implies that no value of $T_1$ and $T_2$ can verify $T_2 > \mathrm{e}^{\mu\theta} T_1$, but allows $T_2 < \mathrm{e}^{\mu\theta} T_1$. For the unequal case, the pulley is in a sticking state and no sliding motion occurs at the current time. In the recent work of [20], the equality constraint (13) is implemented on every single pulley in the sliding cable element, and then a trust-region algorithm [32] is used to essentially determine which sides does the pulley slide towards (or which side has the larger tension). This approach is suitable for the case in which it is ensured that all pulleys have been slid, but it cannot handle the case when there are some sticking pulleys.

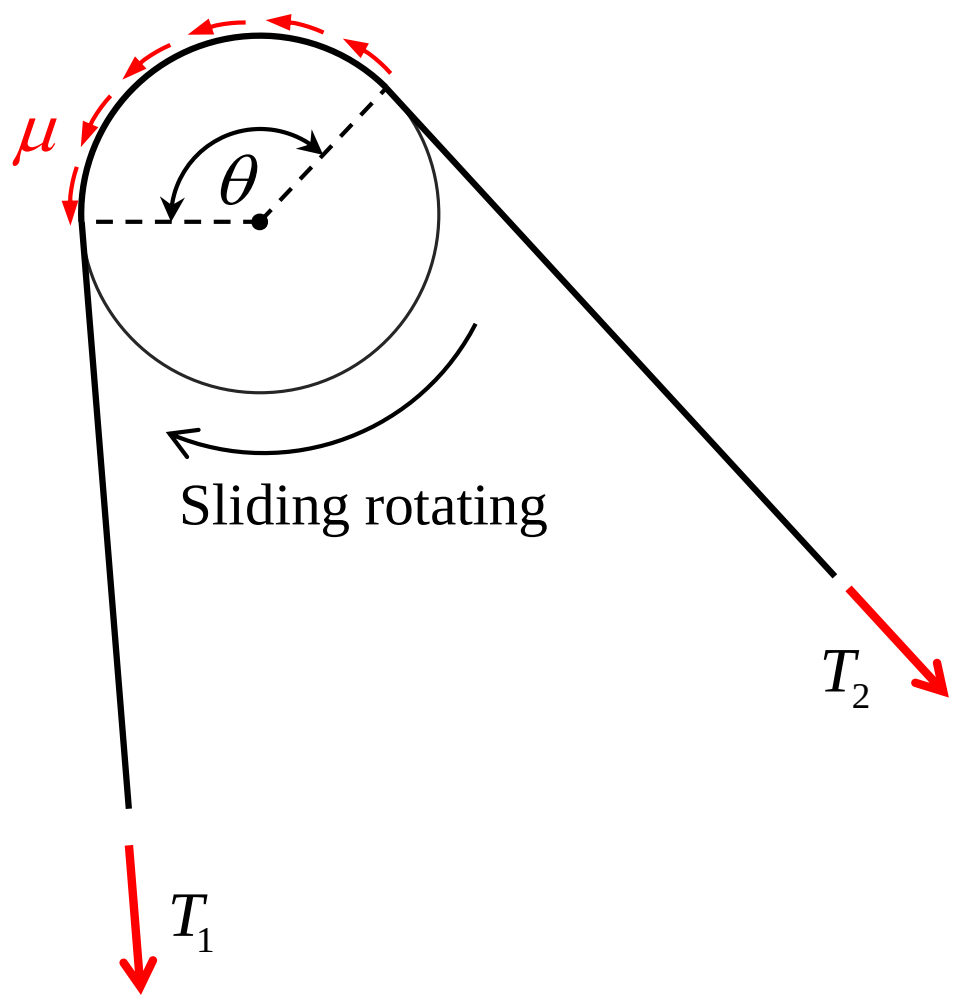

Fig. 4 A cable passing through a pulley.

It is also instructive to gain an insight into sliding effects of the pulley from a dynamically balanced point of view. If at the previous time step, the pulley has been stuck, and at the next time step it is found that $T_2$ is "slightly beyond" $e^{\mu\theta}T_1$, then the pulley cannot provide enough frictions to balance the tension gap between the two sides. The material point of the cable clinging to the pulley will be pulled to the side of the large tension ($T_2$) to meet the capstan equation (13), and as a result, the pulley will, relatively, slide towards the side of the small tension ($T_1$).

By implementing the above sliding criterion on every pulley, and considering that a pulley may slide to either side or even not slide at all (is sticking), we have the following relationship for every pulley.

$$\begin{cases} T_i = T_{i+1}/\alpha_i & s_i > 0 \\ \alpha_i T_{i+1} \le T_i \le T_{i+1}/\alpha_i & s_i = 0 \\ T_i = \alpha_i T_{i+1} & s_i < 0 \end{cases} \tag{14}$$

where $\alpha_i = 1/e^{\mu_i\theta_i}\ (\le 1)$, and $\theta_i$ and $\mu_i\ (\ge 0)$ are, respectively, the total contact angle and the friction coefficient corresponding to the *i*th pulley. The value of $\theta_i$ can be calculated directly from the current configuration of the cable. One cannot generally determine the unknown sliding length variable for every single pulley independently or recursively from expression (14), because the tensions in all the segments are fully coupled. To obtain the solutions, a linear complementarity approach is used, as detailed below.

For each pulley, define the following four intermediate variables:

$$s_i^+ = \frac{|s_i| + s_i}{2} \tag{15}$$

$$s_i^- = \frac{|s_i| - s_i}{2} \tag{16}$$

$$\varsigma_i^+ = T_{i+1}/\alpha_i - T_i \tag{17}$$

$$\varsigma_i^- = T_i - \alpha_i T_{i+1} \tag{18}$$

In Eqs. (15) and (16), the symbol $|(\bullet)|$ means to compute the absolute value of the real number $(\bullet)$. It is thus easy to verify that the sliding length of the pulley can be expressed as:

$$s_i = s_i^+ - s_i^- \tag{19}$$

and the relationship (14) has an equivalence to the following two pairs of complementary relationships:

$$\begin{cases} s_i^+ \varsigma_i^+ = 0; s_i^+ \geq 0; \varsigma_i^+ \geq 0 \\ s_i^- \varsigma_i^- = 0; s_i^- \geq 0; \varsigma_i^- \geq 0 \end{cases} \tag{20}$$

To write the complementary relationships in an element level, by collecting the intermediate variables, the following column vectors are defined.

$$\boldsymbol{s}^+ = \left[ s_1^+, s_2^+, \ldots, s_{n-1}^+ \right]^{\mathrm{T}} \in \Re^{(n-1)} \tag{21}$$

$$\boldsymbol{s}^- = \left[ s_1^-, s_2^-, \ldots, s_{n-1}^- \right]^{\mathrm{T}} \in \Re^{(n-1)} \tag{22}$$

$$\boldsymbol{\varsigma}^+ = \left[ \varsigma_1^+, \varsigma_2^+, \ldots, \varsigma_{n-1}^+ \right] \in \Re^{(n-1)} \tag{23}$$

$$\boldsymbol{\varsigma}^- = \left[ \varsigma_1^-, \varsigma_2^-, \ldots, \varsigma_{n-1}^- \right] \in \Re^{(n-1)} \tag{24}$$

Thus, from Eq. (19), we have

$$\boldsymbol{s} = \boldsymbol{s}^+ - \boldsymbol{s}^- \tag{25}$$

In accordance with (20), the following complementary relationships hold

$$\begin{cases} 0 \leq \boldsymbol{s}^+ \perp \boldsymbol{\varsigma}^+ \geq 0 \\ 0 \leq \boldsymbol{s}^- \perp \boldsymbol{\varsigma}^- \geq 0 \end{cases} \tag{26}$$

Substituting Eqs. (8), (9) and (11) into Eqs. (17) and (18), and writing them in a matrix form, we have the following expressions of $\boldsymbol{\varsigma}^+$ and $\boldsymbol{\varsigma}^-$:

$$\begin{cases} \boldsymbol{\varsigma}^+ = \boldsymbol{M}^+ \boldsymbol{s} + \boldsymbol{q}^+ \\ \boldsymbol{\varsigma}^- = -\boldsymbol{M}^- \boldsymbol{s} + \boldsymbol{q}^- \end{cases} \tag{27}$$

where

$$
\boldsymbol{M}^{+}=\begin{bmatrix} k_1+k_2/\alpha_1 & -k_2/\alpha_1 & & & \\ -k_2 & k_2+k_3/\alpha_2 & -k_3/\alpha_2 & & \\ & \cdots & \cdots & \cdots & \\ & & -k_{n-2} & k_{n-2}+k_{n-1}/\alpha_{n-2} & -k_{n-1}/\alpha_{n-2} \\ & & & -k_{n-1} & k_{n-1}+k_n/\alpha_{n-1} \end{bmatrix} \in \Re^{(n-1)\times(n-1)} \quad (28)
$$

$$
\boldsymbol{M}^{-}=\begin{bmatrix} k_1+\alpha_1 k_2 & -\alpha_1 k_2 & & & \\ -k_2 & k_2+\alpha_2 k_3 & -\alpha_2 k_3 & & \\ & \cdots & \cdots & \cdots & \\ & & -k_{n-2} & k_{n-2}+\alpha_{n-2}k_{n-1} & -\alpha_{n-2}k_{n-1} \\ & & & -k_{n-1} & k_{n-1}+\alpha_{n-1}k_n \end{bmatrix} \in \Re^{(n-1)\times(n-1)} \quad (29)
$$

and

$$
\boldsymbol{q}^{+}=\begin{bmatrix} -k_1\left(l_1-l_{1,0}\right)+k_2\left(l_2-l_{2,0}\right)/\alpha_1 \\ -k_2\left(l_1-l_{2,0}\right)+k_3\left(l_3-l_{3,0}\right)/\alpha_2 \\ \vdots \\ -k_{n-2}\left(l_{n-2}-l_{n-2,0}\right)+k_{n-1}\left(l_{n-1}-l_{n-1,0}\right)/\alpha_{n-2} \\ -k_{n-1}\left(l_{n-1}-l_{n-1,0}\right)+k_n\left(l_n-l_{n,0}\right)/\alpha_{n-1} \end{bmatrix} \in \Re^{(n-1)} \quad (30)
$$

$$
\boldsymbol{q}^{-}=\begin{bmatrix} k_1\left(l_1-l_{1,0}\right)-\alpha_1 k_2\left(l_2-l_{2,0}\right) \\ k_2\left(l_1-l_{2,0}\right)-\alpha_2 k_3\left(l_3-l_{3,0}\right) \\ \vdots \\ k_{n-2}\left(l_{n-2}-l_{n-2,0}\right)-\alpha_{n-2}k_{n-1}\left(l_{n-1}-l_{n-1,0}\right) \\ k_{n-1}\left(l_{n-1}-l_{n-1,0}\right)-\alpha_{n-1}k_n\left(l_n-l_{n,0}\right) \end{bmatrix} \in \Re^{(n-1)} \quad (31)
$$

The above expressions of $\boldsymbol{\varsigma}^{+}$ and $\boldsymbol{\varsigma}^{-}$ are a set of *linear* equations purely in terms of unknowns $\boldsymbol{s}_i$. Substituting Eq. (25) into Eq. (27) yields

$$
\begin{bmatrix} \boldsymbol{\varsigma}^{+} \\ \boldsymbol{\varsigma}^{-} \end{bmatrix} = \begin{bmatrix} \boldsymbol{M}^{+} & -\boldsymbol{M}^{+} \\ -\boldsymbol{M}^{-} & \boldsymbol{M}^{-} \end{bmatrix} \begin{bmatrix} \boldsymbol{s}^{+} \\ \boldsymbol{s}^{-} \end{bmatrix} + \begin{bmatrix} \boldsymbol{q}^{+} \\ \boldsymbol{q}^{-} \end{bmatrix} \quad (32)
$$

By defining $\boldsymbol{z}=\begin{bmatrix} \boldsymbol{s}^{+} \\ \boldsymbol{s}^{-} \end{bmatrix} \in \Re^{2(n-1)}$, $\boldsymbol{q}=\begin{bmatrix} \boldsymbol{q}^{+} \\ \boldsymbol{q}^{-} \end{bmatrix} \in \Re^{2(n-1)}$, $\boldsymbol{w}=\begin{bmatrix} \boldsymbol{\varsigma}^{+} \\ \boldsymbol{\varsigma}^{-} \end{bmatrix} \in \Re^{2(n-1)}$ and

$$
\boldsymbol{M}=\begin{bmatrix} \boldsymbol{M}^{+} & -\boldsymbol{M}^{+} \\ -\boldsymbol{M}^{-} & \boldsymbol{M}^{-} \end{bmatrix} \in \Re^{2(n-1)\times 2(n-1)} \quad (33)
$$

we have the final LCP for the analysis of the current time step:

$$
\begin{cases} \boldsymbol{w}=\boldsymbol{M}\boldsymbol{z}+\boldsymbol{q} \\ \boldsymbol{0}\le \boldsymbol{w} \perp \boldsymbol{z} \ge \boldsymbol{0} \end{cases} \quad (34)
$$

Once a solution of the above LCP is obtained, sliding lengths of all the pulleys can be directly obtained by using Eq. (25). Substituting these sliding length values into Eq. (12) yields the final nodal force vectors of the sliding cable element. Very importantly, the *non-loaded* length of each segment should be adjusted according to the sliding lengths of pulleys (Eq. (8)) and saved for the analysis of next time step. A particular case is worth to notice: if $\boldsymbol{q} \geq 0$, one can immediately find a solution of such LCP: $\boldsymbol{s}^{+} = \boldsymbol{s}^{-} = \boldsymbol{0} \left(= \boldsymbol{s}\right)$, demonstrating in this case that all the pulleys are sticking; by a priori employment of $\boldsymbol{s} = \boldsymbol{0}$ and in combination with Eqs. (8), (9) and (11), one can easily verify that $\boldsymbol{q} \geq 0$ is exactly compatible with the inequality constraint (second expression) of the sliding criterion (14). This, on a side note, demonstrates the correctness of the proposed approach for such a particular case.

**Remark 1**: By taking advantage of the small deformation assumption and using the approximate relation Eq. (10), the tensions, and eventually $\boldsymbol{w}$ ($\varsigma^{+}$ and $\varsigma^{-}$) are written mathematically as a *linear* function of the sliding distance $s_i$. This is important for the above implementation of linear complementarity approach, otherwise nonlinear complementarity relationships will be obtained which will largely complicate the analysis.

**Remark 2**: We have concluded previously that analyses involving sliding cables with frictions are dependent of the loading path. This can be explained more clearly from the mathematical formulation of LCP (34): the historical information $l_{i,0}$ is embraced in expressions of $q$ ($\boldsymbol{q}^{+}$ and $\boldsymbol{q}^{-}$), and thus it has an influence to solutions of sliding lengths, as well as the final nodal force vectors.

3.3 Discussions of the solutions and extension to implicit analyses.

The above explanations are mainly given from a numerical point of view, by assuming that the solution of the LCP can be found. On the other hand, from a theoretical point of view, naturally these questions arise: does LCP (34) always has a solution; if it has a solution, is such solution unique? For the former question, unfortunately, it is unable to give a complete conclusion currently. But there are still some instructive conclusions worthy of mention. Due to the equivalence of complementary relationship (20) and the relationship defined directly by sliding criterion (14), if LCP (34) has no solution, then any of existing predictor-corrector-based approaches cannot find a solution, and the problem itself is possibly non-physical. By our preliminary numerical tests, it seems that this case occurs when the whole sliding cable is slack (the length of the whole cable after deformation is less than its *non-loaded* length). A glance into this unsolvable case may be given from the second expression of Eq. (14), which is a contradiction when $T < 0$. However, is cable slacking represents the sufficient or necessary condition for the unsolvable characteristic of LCP (34)? This question remains open and deserves further investigation. While for the latter question, by analyzing the property of the coefficient matrix $\boldsymbol{M}$, we will give a complete conclusion in what follows.

Note that the expressions $\alpha_i \geq 0$ $(1 \leq i \leq n-1)$ and $k_i > 0$ $(1 \leq i \leq n)$ always hold. By using induction, it is can be first found that all principal minors of the tri-diagonal matrices $\boldsymbol{M}^+$ and $\boldsymbol{M}^-$ are positive (and obviously they are invertible), and for any index set $\alpha$ for $\boldsymbol{M}$, if there exists at least one index $\gamma(\leq n-1)$ such that both $\gamma \in \alpha$ and $n-1+\gamma \in \alpha$, then $\det \boldsymbol{M}_{\alpha\alpha} = 0$, otherwise $\det \boldsymbol{M}_{\alpha\alpha} > 0$. For such an index $\gamma$, it is straightforward to see from the structure of matrix $\boldsymbol{M}$ (Eq. (33)) that

$$\boldsymbol{M}(:,\gamma) = \begin{bmatrix} \boldsymbol{M}^+(:,\gamma) \\ -\boldsymbol{M}^-(:,\gamma) \end{bmatrix} \in \Re^{(n-1)} \quad \text{and} \quad \boldsymbol{M}(:n-1+\gamma) = \begin{bmatrix} -\boldsymbol{M}^+(:,\gamma) \\ \boldsymbol{M}^-(:,\gamma) \end{bmatrix} \in \Re^{(n-1)}, \quad \text{thus}$$

$\boldsymbol{M}(:,\gamma) + \boldsymbol{M}(:n-1+\gamma) = \boldsymbol{0} \in \Re^{(n-1)}$, demonstrating these two column vectors of $\boldsymbol{M}$ are linearly dependent. Using the theorem presented in Section 2.2, the LCP (34) is ***w***-unique, i.e., $\varsigma^+$ and $\varsigma^-$ are unique for any solutions of the LCP. As $\boldsymbol{M}^+$ and $\boldsymbol{M}^-$ are invertible, by using any part of Eq. (27), one can conclude that the sliding length vector $\boldsymbol{s}$ (but not necessarily $\boldsymbol{s}^+$ & $\boldsymbol{s}^-$) is unique. This completes the proof.

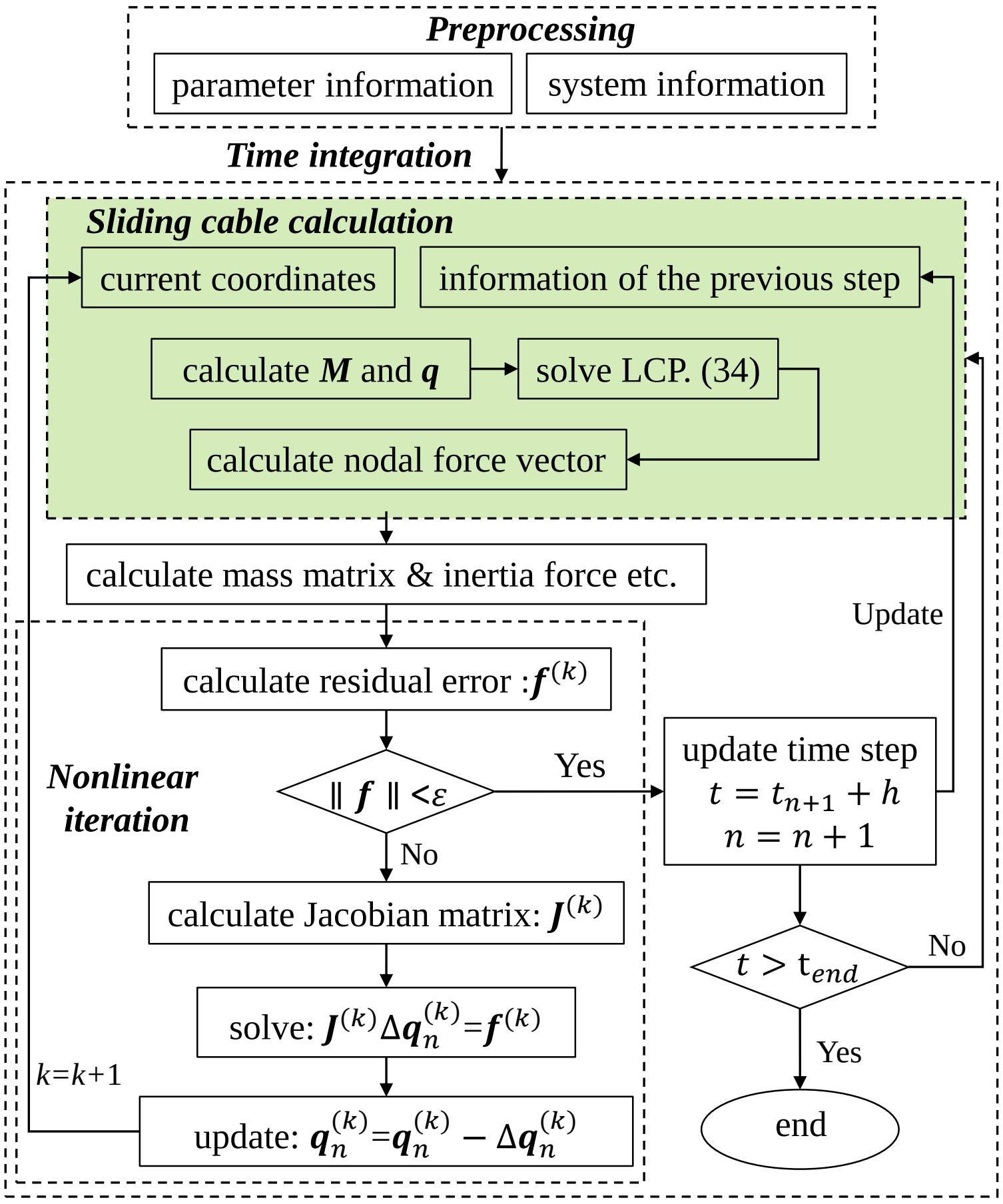

Fig. 5 Computational chart for implicit dynamic analysis with sliding cables.

For the sake of clarity, all the above analysis presented in this section is written in the context of an explicit dynamic analysis. Extensions to implicit dynamic or static analyses are straightforward. For an implicit dynamic analysis, the discrete dynamic equilibrium equation is formulated at each time step, which commonly yields a set of nonlinear equations. For the current time step, started by a given estimated solution, an iterative approach may be used to obtain the accurate solution. Within each iteration, the core task is exactly to evaluate the unbalance force vector under the current updated solution. For a structure containing sliding cables, the nodal internal force vector contributed by each sliding cable can be calculated using the approach proposed above. A brief computational chart is shown in Fig. 5. Some details are worth noting. As the analysis is dependent of the loading path, one must select a reference historical solution and extract the *non-loaded* length information of each segment ($l_{i,0}$) to formulate the LCP (or more narrowly, $\boldsymbol{q}$). For all iterations in the current time step, the reference solution should remain as the solution of the *previous time step*, rather than the solution of the *previous iterative step*. In the (iteration-based) static analysis, an essentially similar process is conducted. For a static analysis using a multi-step loading strategy, the reference solution used in every loading step should correspond to the previous loading step; for a one-step loading strategy, such a reference solution naturally corresponds to the initial unloaded state. Another important issue involved in the implicit analysis is the deduction of stiffness matrix, which can be commonly obtained by taking the derivation of the generalized force vector (the nodal force vector, here) with respect to the generalized coordinates (the nodal position vector, here). However, for sliding cable analysis with frictions, the exact stiffness matrix relevant to the sliding cable element cannot be given analytically; this is because of the unpredictability of the motion states of pulleys. In other word, the sliding length vector $\boldsymbol{s}$ governed by LCP (34) cannot be expressed explicitly and smoothly as a function such as $\boldsymbol{s}(\boldsymbol{q}_{\mathrm{s}})$, and the solution of $\boldsymbol{s}$ may be non-smooth with variations of $\boldsymbol{q}_{\mathrm{s}}$, which eventually leads to an non-differentiable characteristic of the nodal force vector at some points. For implicit *dynamic* analysis in combination with Newton-Raphson method, if a small time step size is adopted, the stiffness matrix of the sliding cables has only a little contribution to the Jacobian matrix in each iteration (the Jacobian matrix is dominated by mass matrix) and thus it can be neglected. For implicit *static* analysis, due to the absence of mass matrix, the stiffness matrix dominates the Jacobian matrix. In this case, an approximate stiffness matrix of sliding cable elements can be tentatively evaluated by using a numerical difference technique, which is adopted in our following numerical examples. A more effective way for static analysis might be in combination with the so called dynamic relaxation approach [9, 20, 33], which is essentially equivalent to the explicit dynamic analysis presented previously and is Jacobian free. Exhaustive discussions on efficiency of various kinds of approach in solving the *system* governing equation is not of interest in this paper.

## 4. Numerical examples

Three numerical examples are presented in this section to demonstrate the effectiveness of the proposed approach, as well as to reveal some novel phenomenon involved in sliding cables with frictions. For all the analyses, the relevant LCPs are solved by the popular Lemke's algorithm.

### 4.1 A sliding cable passing through two fixed pulleys

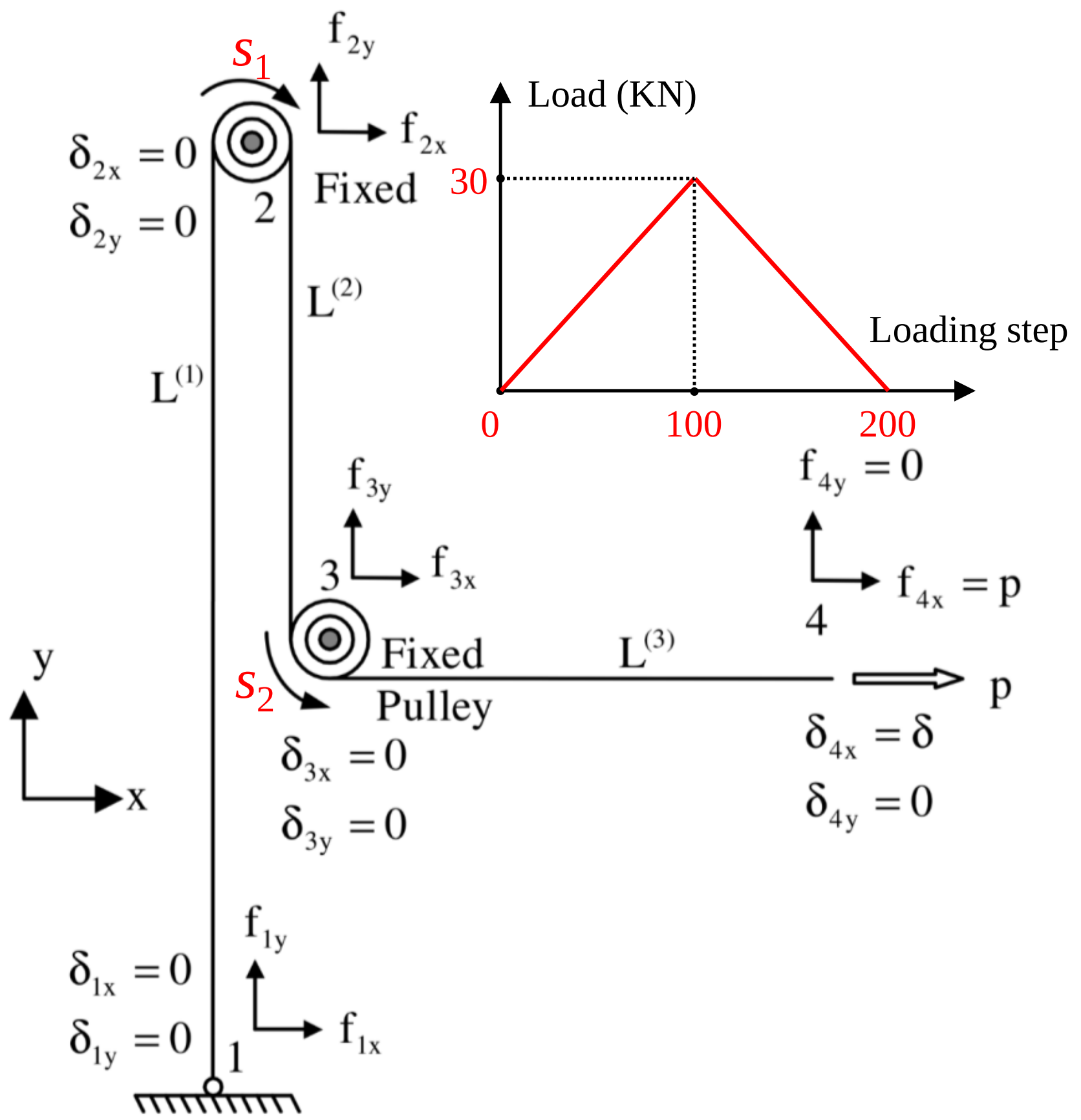

Fig. 6 A sliding cable passing through two fixed pulleys (taken from [1] and slightly modified).

The first example concerns with static analyses of a sliding cable passing through two fixed pulleys, which was first presented in [1] and recently investigated in [20]. As shown in Fig. 6, the cable is fixed at node 1 and can slide at two fixed pulleys. The pulleys are assumed sufficiently small and can be modeled as two nodes 2 and 3. All the material and dimensional parameters are identical with the ones in [20]. The cable is assumed linear elastic, weightless and undamped, with a total *non-loaded* length of $L$ = 240 cm and a cross section area of $A$ = 0.6 cm$^2$. It is made of stainless-steel with an elastic modulus of 115 GPa. As indicated in Fig. 6, the cable is divided into three segments: $L^{(1)}$, $L^{(2)}$ and $L^{(3)}$ having a length of 100 cm, 40 cm and 100 cm, respectively. A unified friction coefficient $\mu$ = 0.05 is considered for both pulleys, and the total contact angles are taken as $\theta_1 = \pi$ at node 2 and $\theta_2 = \pi/2$ at node 3. A pulling force $P$ with a peak value of 30 KN is applied to the end node of the cable. A major difference from the existing works [1, 20] is that we consider here a whole process of loading-unloading: the pulling force is first loaded and then unloaded, both via a multistep strategy (200 steps in total), as clearly indicated in the upper right of Fig. 6. Instead of using a one-step loading strategy, this is particularly introduced to investigate the holistic behavior of cable sliding effects. Static loading is assumed and the analyses are carried out by using Newton-Raphson method, with a convergence error $\xi$ = 1e-7.

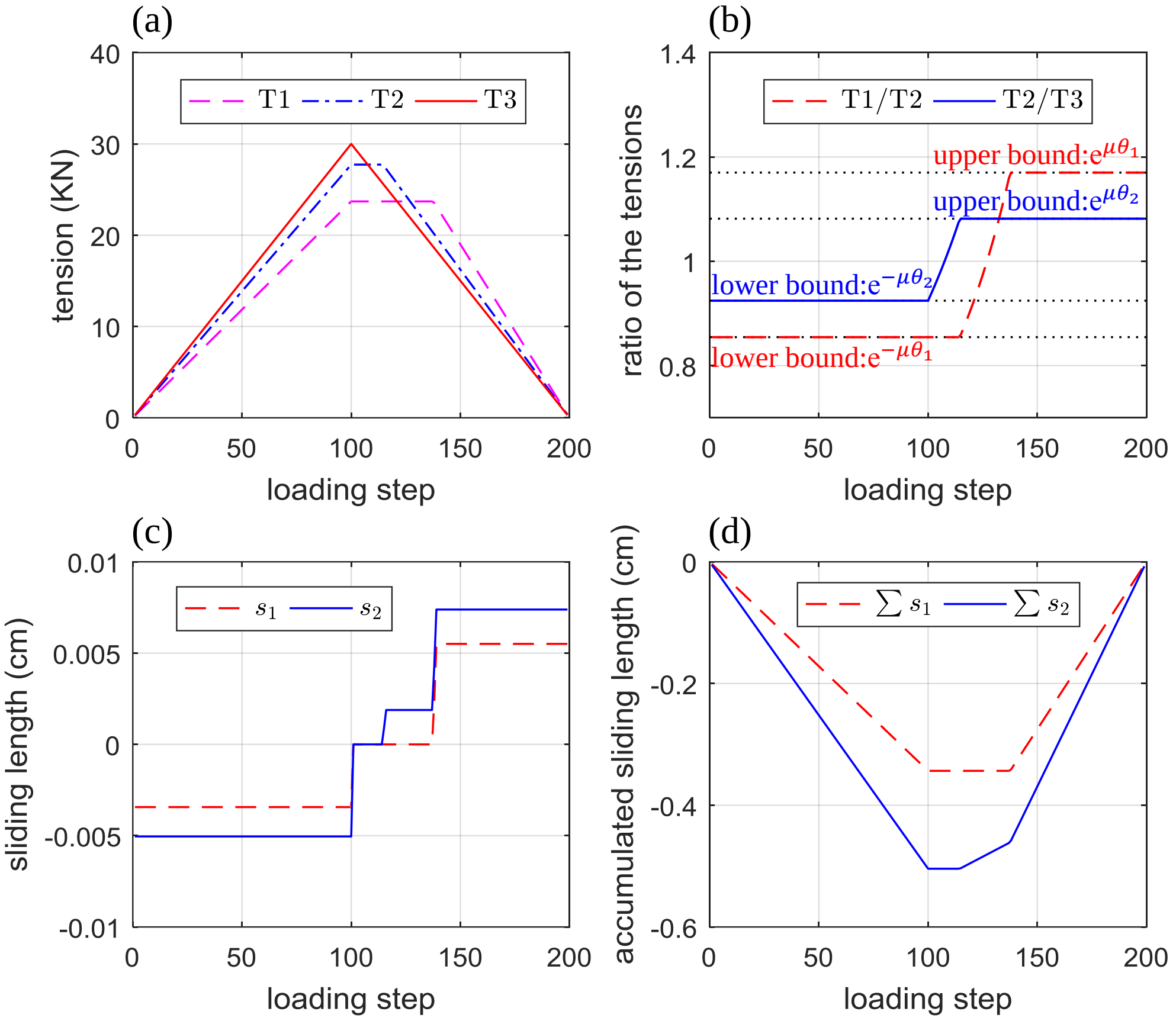

Fig. 7 Some representative results vs. loading step: (a) tensions; (b) tension ratios; (c) one-step sliding lengths; (d) accumulated sliding lengths.

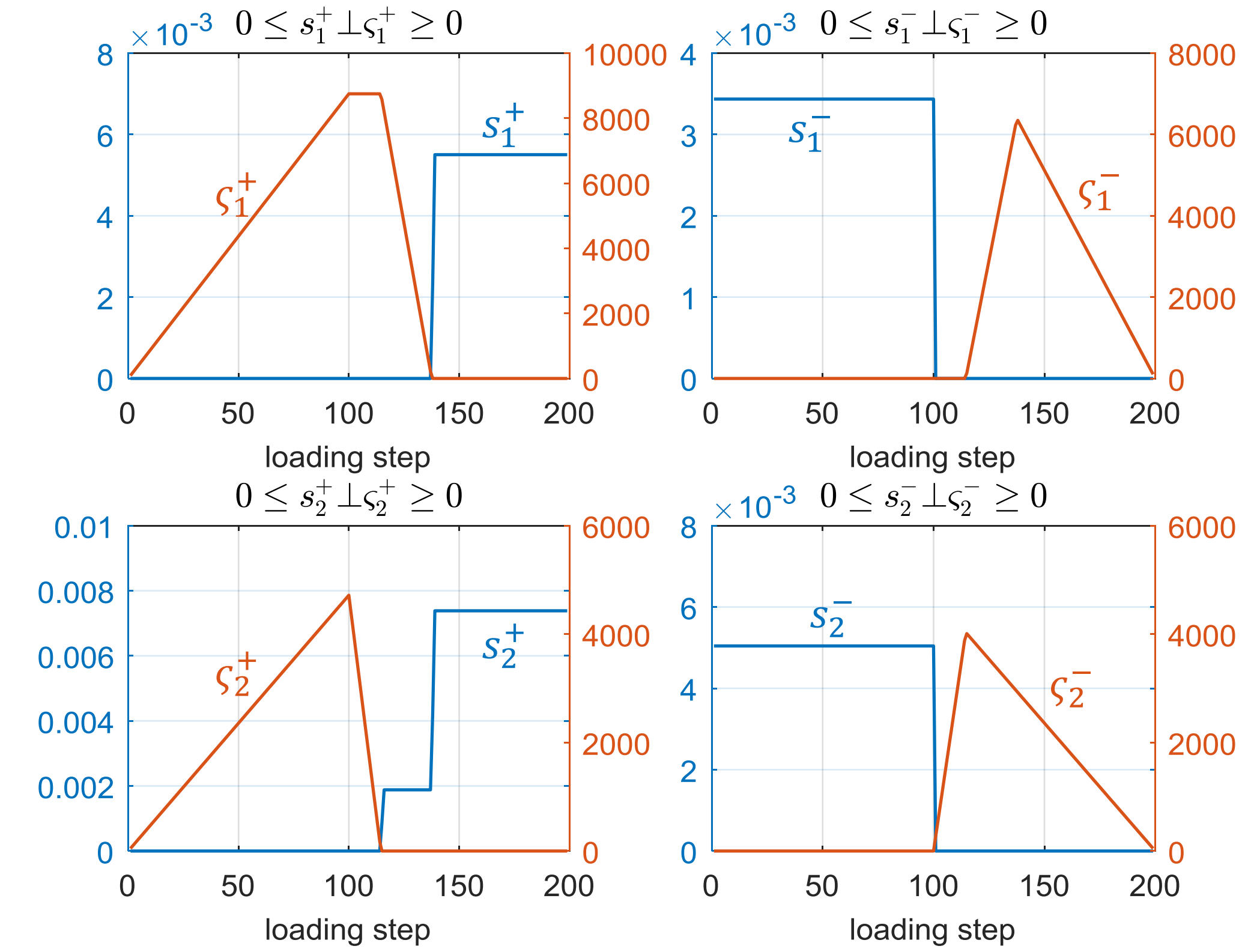

Fig. 8 Complementary relationships for the four pairs of intermediate variables.

Fig. 7 (a) and (b) give tensions in the three segments and their ratios, with evolutions of loading step; Fig. 7 (c) and (d) give (one-step) sliding lengths and accumulated sliding lengths of the two pulleys. These figures bring insight into the whole loading-unloading process. During the loading phase, all the tensions increase monotonously. The two pulleys are both sliding towards node 1, leading to monotonous increases of the absolute values of their accumulated sliding lengths. The tension ratios are coincide with "lower bounds" defined by sliding criterion. While in the unloading phase, the results are much complicated. Motion states of the two pulleys are not always synchronous. In the beginning of the unloading phase, the pulleys are both sticking, as their sliding lengths are both zero; tension ratios relevant to pulleys both increase until the second pulley first achieves to its corresponding "upper bound", which indicates that it begins to slide. Soon afterwards, tension ratio of the first pulley achieves to its "upper bound" as well, and both pulleys have been sliding until the unloading phase is terminated. For a same force size $P\left(0<P<30\right)$ applied in the two different (loading and unloading) phases, the tension results are completely different, which demonstrates the physical problem is dependent of the loading path. Thus, to accurately analyze sliding cable systems with frictions, one must take into account the loading history. Further calculations show that the total deformation of the cable is less than 1 cm leading to a strain less than 0.5%, which demonstrates the validity of using the small deformation assumption. It is clear to see from Fig. 7 (a) that the tension in the third segment is completely identical with the loading force, which is as expected due to boundary equilibrium of node 4. At the 100th loading step (corresponding to the peak value of the pulling force), tensions in the three segments are 23.7, 27.7, 30.0 KN, and the accumulated sliding lengths of the pulleys are 0.34 and 0.50 cm, respectively. These values agree well with solutions calculated using the analysis formula presented in [1] as well as the numerical solutions directly presented in [20]. In addition, Fig. 8 presents evolution curves of the four pairs of intermediate variables. It can be clearly found that complementary relationships are well satisfied. These facts eventually demonstrate the correctness of the proposed approach.

### 4.2 A single pulley sliding on a continuous cable

The second example concerns with dynamic analyses of a single pulley moving on a cable. A similar system has been presented in [4], where only a frictionless sliding motion is considered. To investigate friction effects, Coulibaly et al [6] recently extended such system to a two-node symmetric sliding system. However, they simply analyzed the dynamic responses in a specified time interval such that the sliding directions (sides) of the nodes are consistent and are known in advance, which avoided facing the challenging issue of sliding state transformation of the nodes. The example presented here exactly put emphases on such a transformation and aims to gain an insight into how it will influence the mechanical responses of the system.

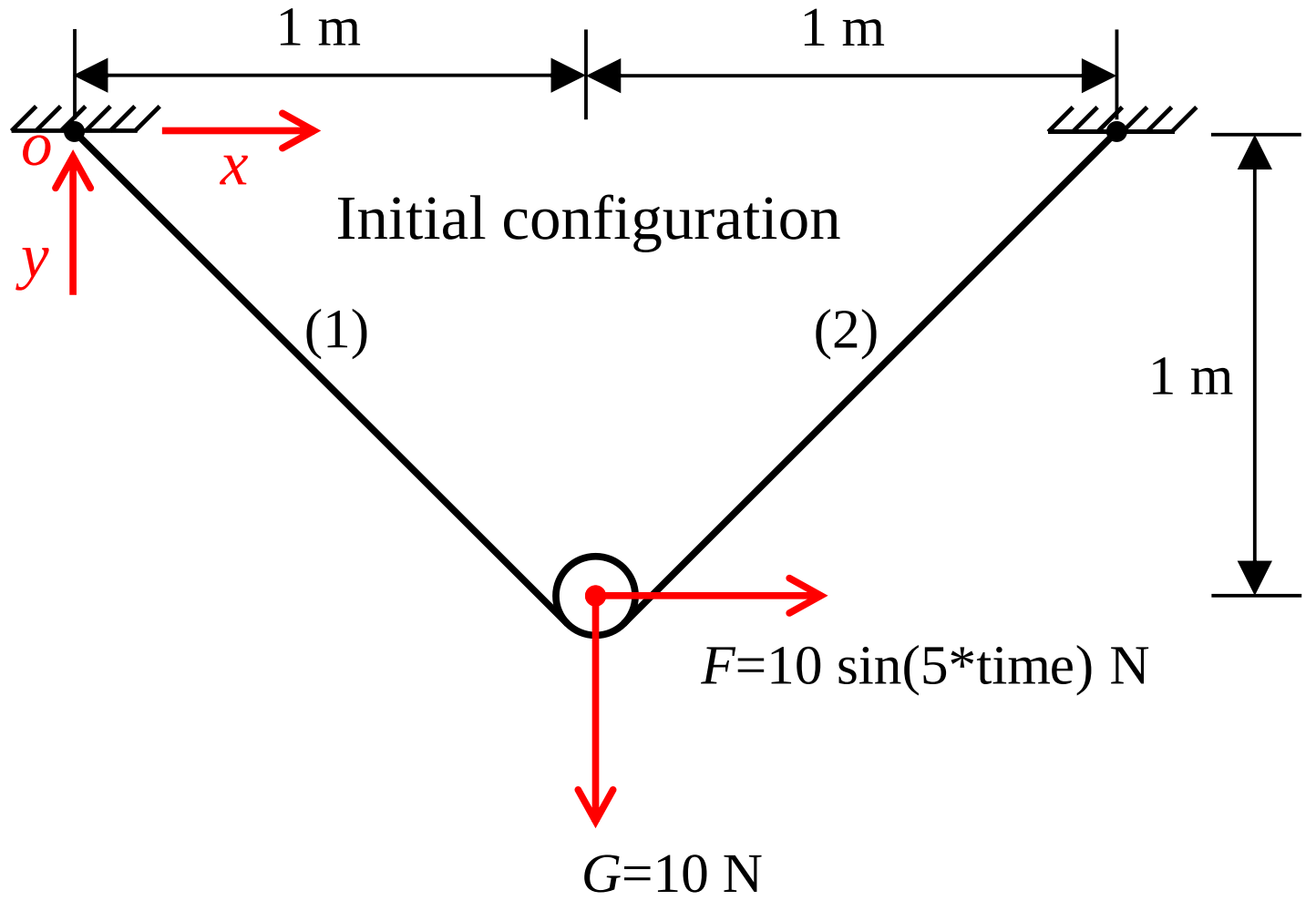

Fig. 9 A pulley moving on a cable subjected to an alternating load.

As shown in Fig. 9, a small pulley with a gravity of $G$ = 10 N is suspended on a massless cable. *oxy* denotes the global coordinate system. Initially, the system is at the static equilibrium state: the pulley is located at ($x_0$ = 1 m, $y_0$ = -1 m), tensions in the two segments are equal to balance the gravity. Then, a horizontal alternating load is applied to the pulley to drive its motion. The cable has an initial length of $L = 2\sqrt{2}$ m with a tensile stiffness denoted as $EA$. The friction coefficient is denoted as $\mu$, and the contact angle $\theta$ is calculated from the current configuration in real time. Dynamic analyses are carried out using the Newmark algorithm with parameters $\alpha$ = 0.3, $\delta$ = 0.5 (symbolic definitions are coincide with [34]). The total analysis time is set as 4 s such that the applied force is able to experience several cycles. The time step size is chosen as $\Delta t$ = 0.0002 s, and the convergence error for Newton-Raphson method in each time step size is $\xi$ = 1e-7.

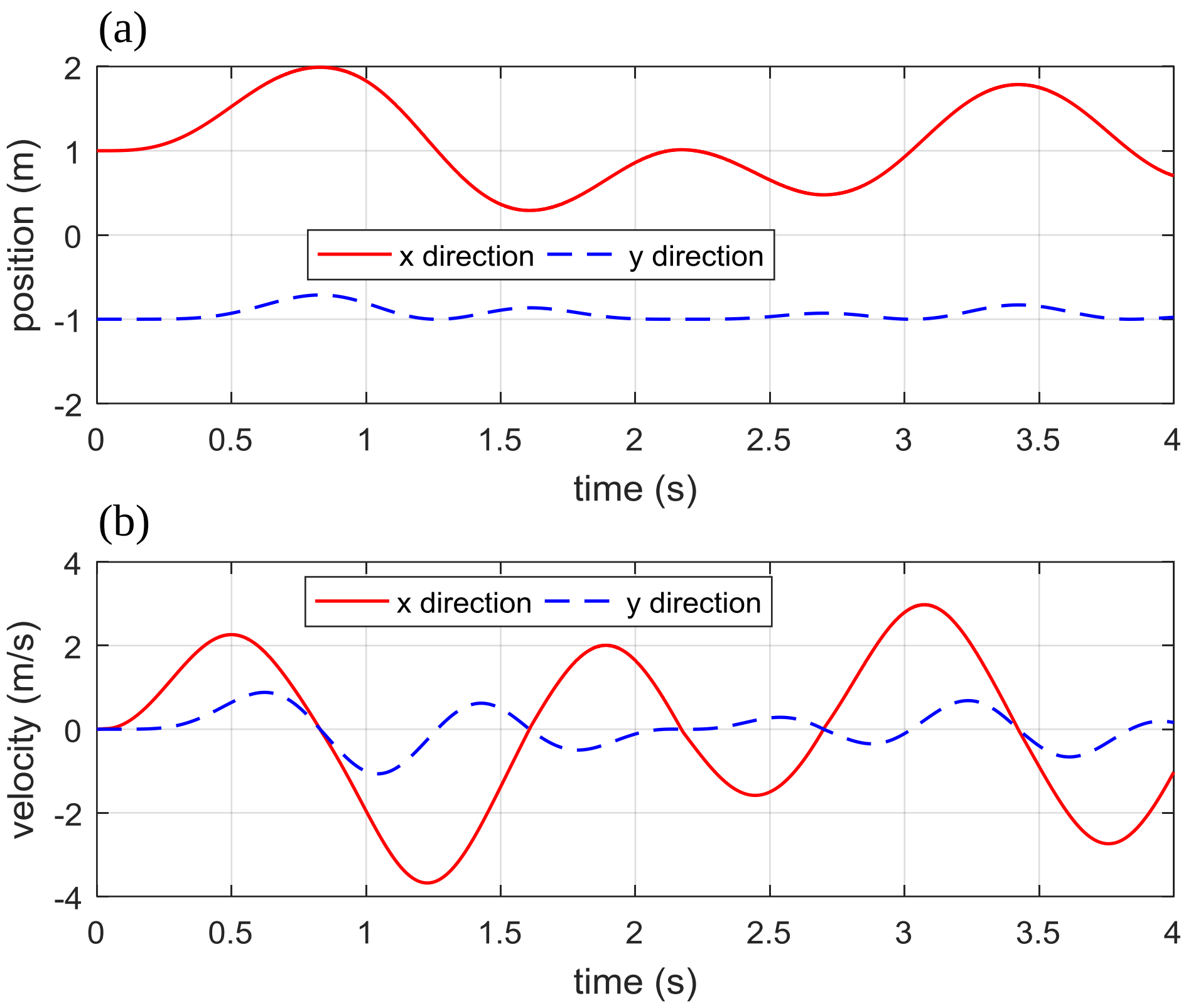

Fig. 10 Time history curves of the (a) position and (b) velocity of the pulley.

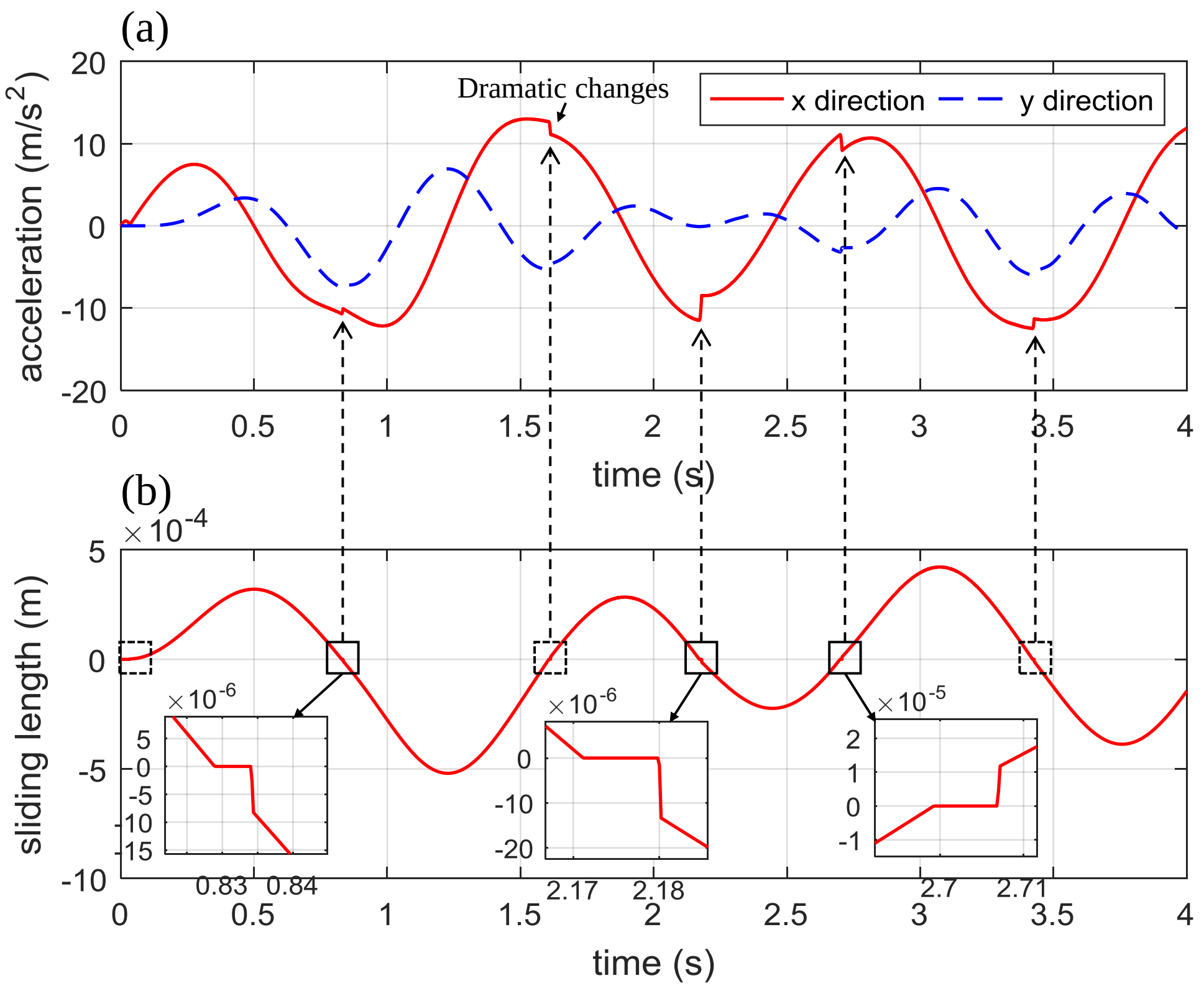

Fig. 11 Time history curves of the (a) acceleration and (b) sliding length of the pulley.

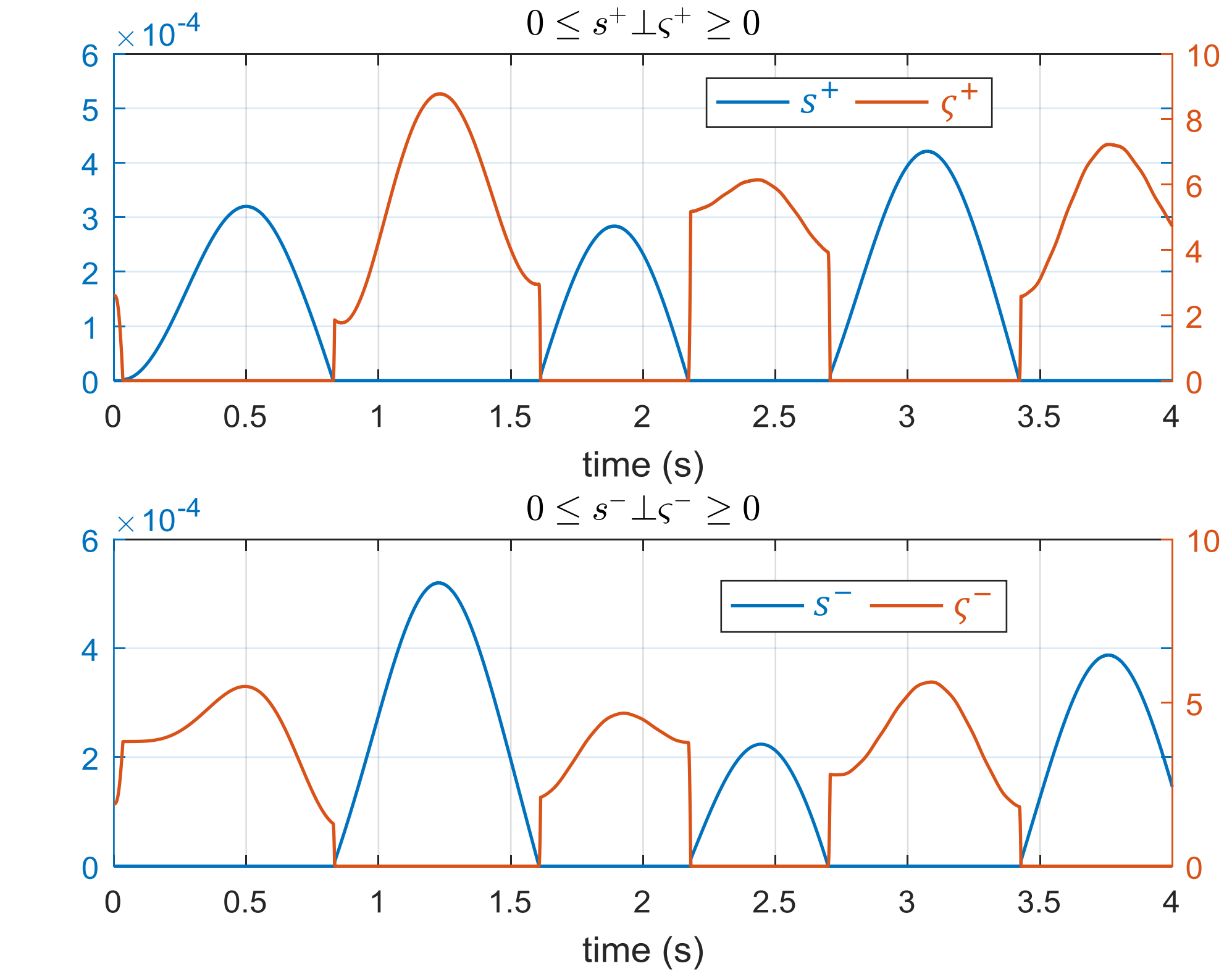

Fig. 12 Time history curves of the two pairs of intermediate variables.

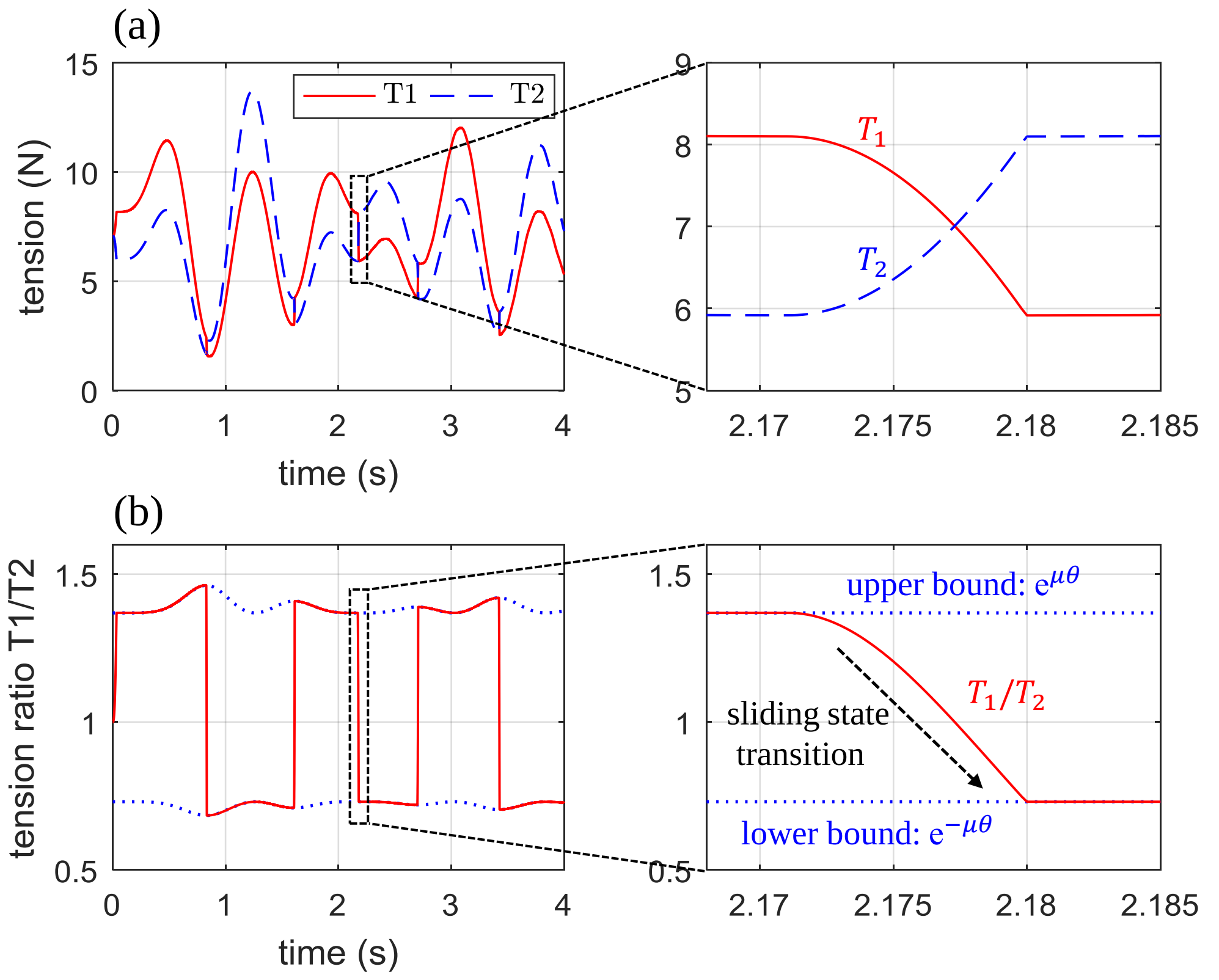

Fig. 13 Time history curves of the (a) tensions and (b) tension ratios, including their enlarged drawings.

We first focus on investigating the system general behaviors under a given friction coefficient $\mu$ = 0.2 and tensile stiffness $EA$ = 1e4 N. Fig. 10 gives time history curves of the position and velocity of the pulley in two directions. Theses curves clearly show a large overall motion of the pulley; the pulley mainly undergoes a horizontal reciprocating motion caused by the alternating load. Trivially, these curves are quite smooth. However, this is not the case for time history curves of acceleration presented in Fig. 11 (a), in which dramatic changes appear at some time points. To further uncover such curious phenomenon, Fig. 11 (b) gives time history curve of the (one-step) sliding length of the pulley. It is clear to see that the pulley is sliding almost all the time (as $s \neq 0$), towards either side ($s$ > 0 or $s$ < 0). Nevertheless, there are still some short-lived periods in which the sliding length is zero, demonstrating in these time periods the pulley is sticking. Actually, these time periods typically mark transitions of the pulley sliding direction. Compared with Fig. 11 (a), it is straightforward to find that these time periods exactly correspond to the above mentioned dramatic changes of acceleration curves. To gain an insight, Fig. 12 gives evolution curves of two pairs of intermediate variables, and clearly, complementary relationships are well satisfied. Fig. 13 (a) and (b) present time history curves of tensions in the two segments and its ratio. Since the pulley is sliding almost all the time, the tension ratio curve exactly coincides with the upper bound or the lower bound according to the sliding criterion. These two bounds are not constant as the contact angle is variable, and different bounds exactly reflect different sliding directions. In the short-lived time periods corresponding to the sticking state of the pulley, the tension ratio is able to reverse (from the upper bound to the lower bound, or conversely), giving rise to some dramatic changes of the tensions, and eventually the accelerations.

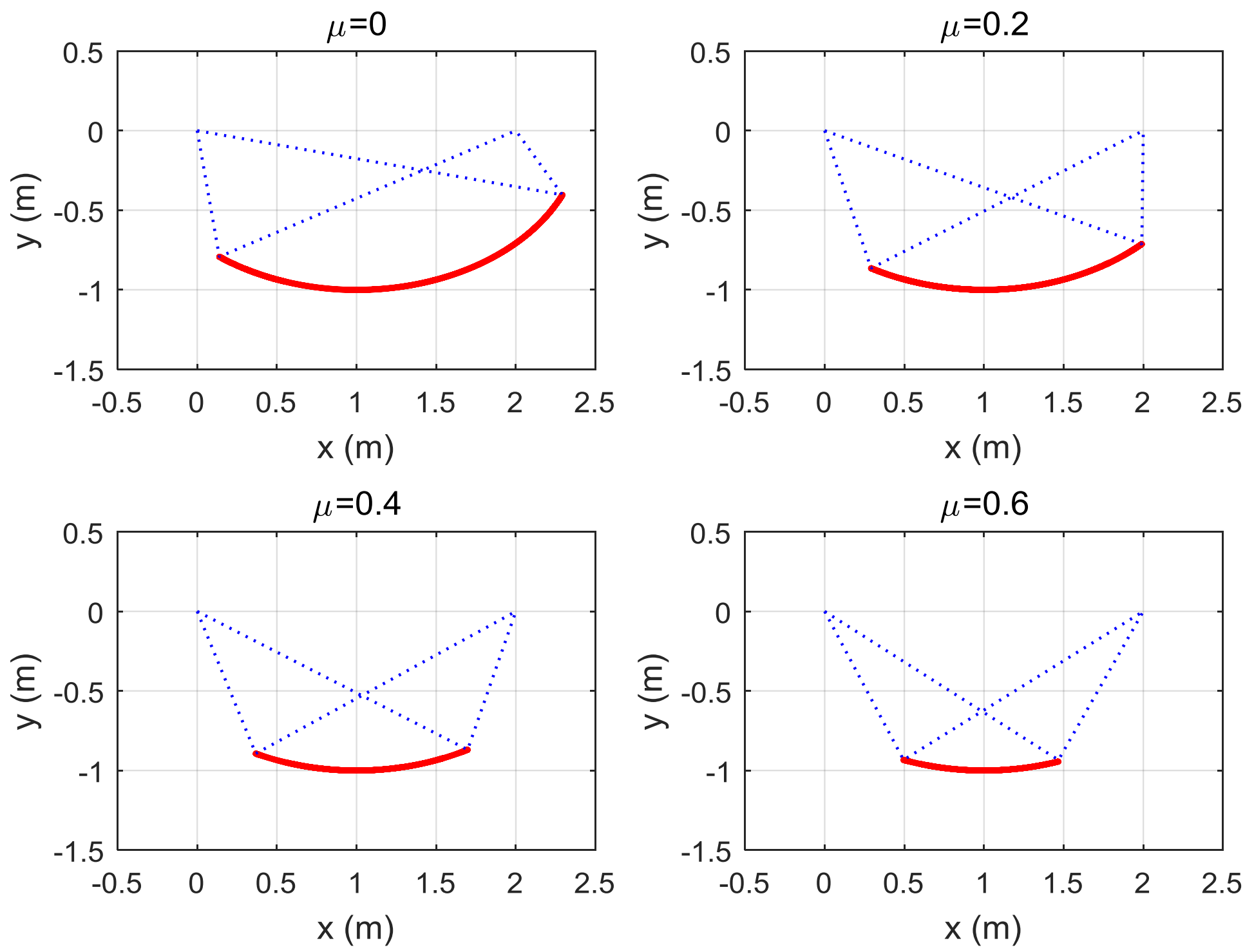

Fig. 14 Trajectory (denoted as red lines) of the pulley for different cases of friction

coefficient.

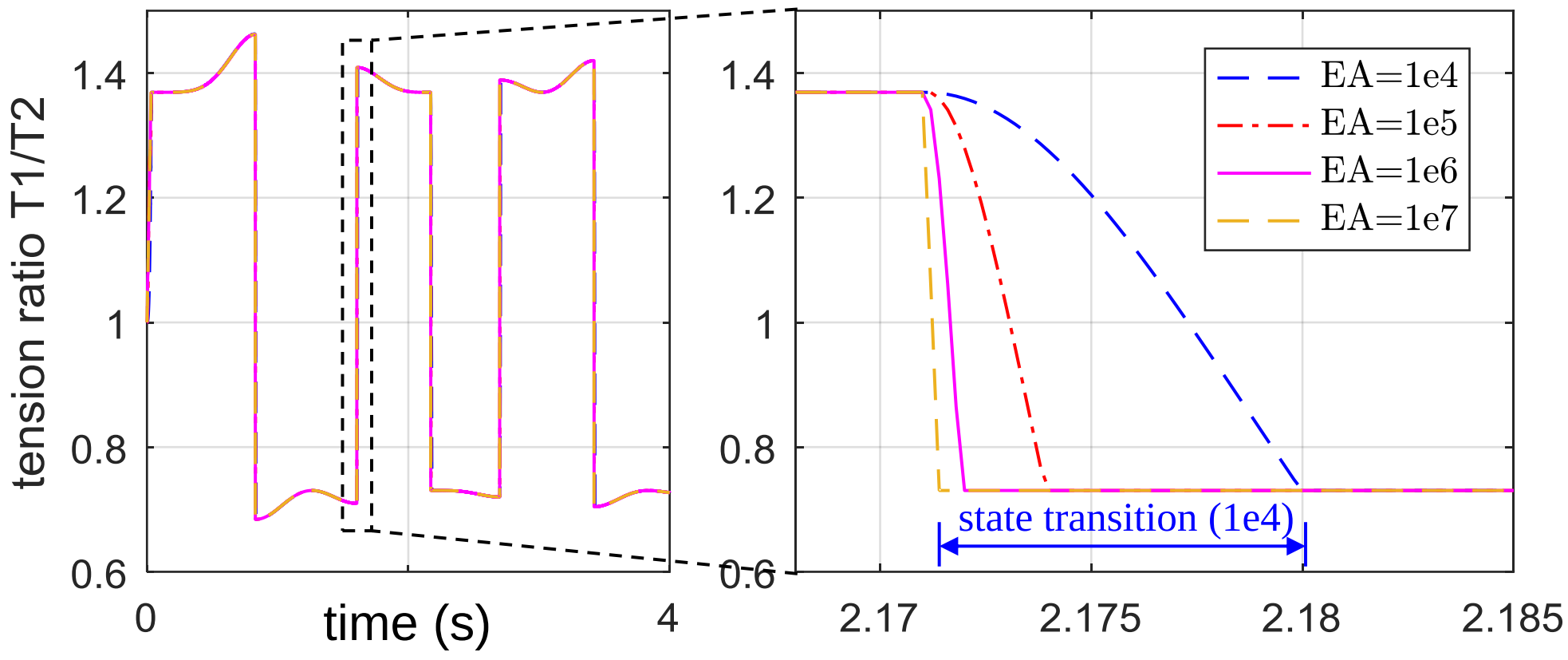

Fig. 15 Time history curves of tension ratios for different cases of tensile stiffness.

The above analyses focus on a dynamic analysis using fixed parameters, with the aim to explore general behaviors of such system. We are also interested in investigating the system behavior under variable parameters. By conducting more numerical analyses, we find that two additional conclusions are worth mentioning. (1) As shown in Fig. 14, the friction coefficient has a significant influence on the range of motion (trajectory) of the pulley: the larger the friction coefficient, the smaller the range of motion. This is not surprising, as frictions typically prevent relative motions in physics. (2) If sufficient large, the tensile stiffness of the cable has little influence on the overall motion of the pulley. However, it has a very closely relationship with the time duration of sliding state transitions: the larger the tensile stiffness, the shorter the time duration. Fig. 15 gives tension ratio curves using different values of the tensile stiffness clearly demonstrating this point. This may explain why in [6] an cable inextensible hypothesis is adopted, because it ensures that when released, the sliding node can immediately slide downwards without any sticking state at the very beginning.

In summary, this simple example highlights the complicated nonlinear mechanical behaviors involved in sliding cable with frictions, particularly induced by sliding state transitions of the pulley. Accurate analyses of such type of system heavily depend on the precise determination of the motion state of the pulley, which can be handled simply and effectively using the proposed linear complementarity approach.

4.3 Folding analysis of a ten-stage clustered tensegrity tower

Previous two examples concern with simple systems purely composed of sliding cable element. In this subsection, a ten-stage clustered tensegrity tower is thoroughly investigated to demonstrate abilities of the proposed approach in a more complicated structure, as well as to reveal some important influences of frictions on the deployment performance of such kind of structure. Analyses of such structure in the frictionless case have been conducted in our previous work [12]. This structure can also be considered as an extension of the lower stage towers studied, mainly from a static aspect, in existing literature [9, 11]. Both static analyses and dynamic analyses are involved in this work.

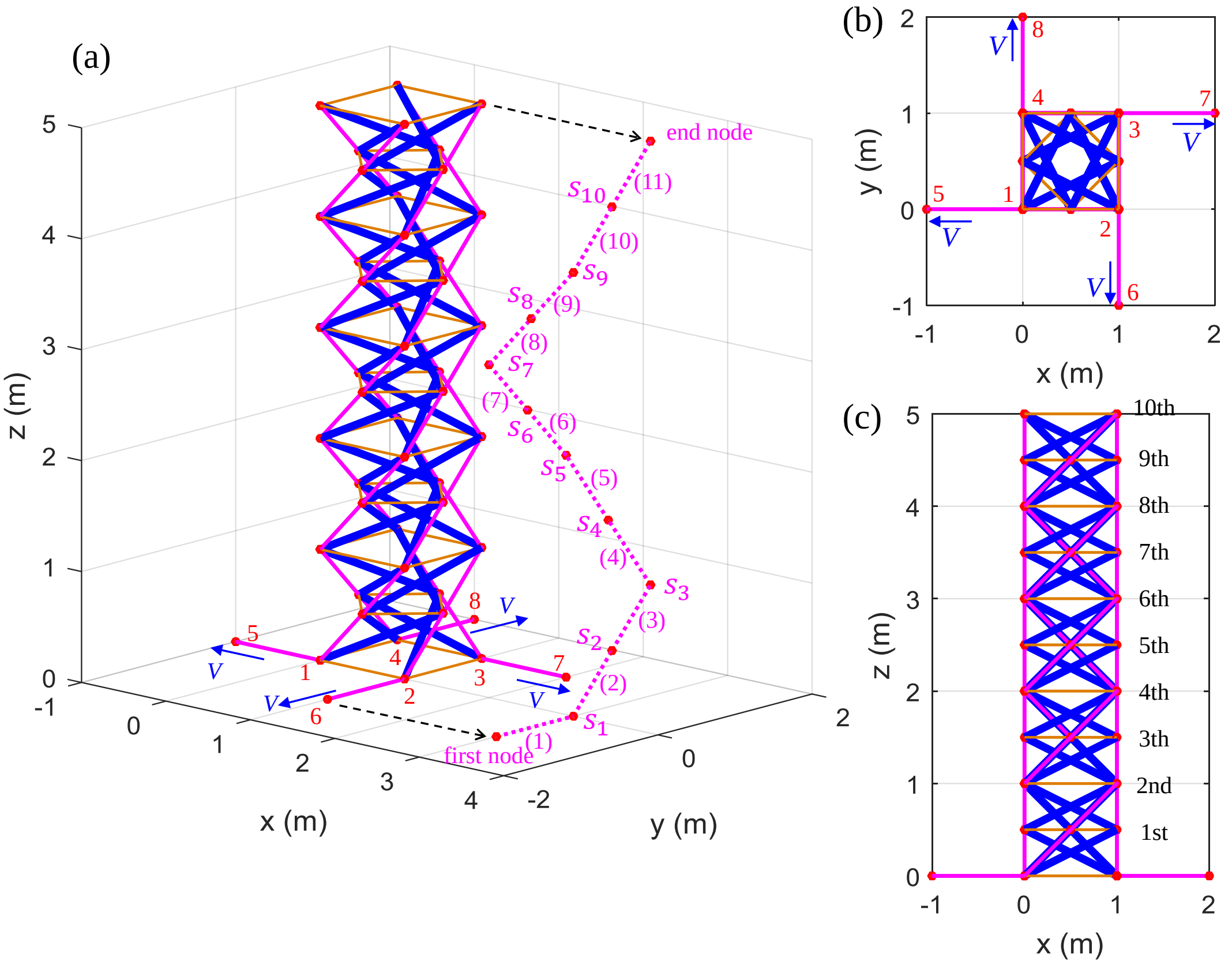

Fig. 16 The ten-stage clustered tensegrity tower: (a) a perspective view, (b) a top view, (c) a side view.

Fig. 16 presents a perspective, a top and a side view of the structure, in which blue lines represent struts, brown lines represent classical cables and pink lines represent sliding cables. The tower is roughly assembled by ten identical quadruplex units, in such a way that 36 strut-to-strut connections occur, which makes it a class 2 [35] tensegrity. Forty original vertical cables are clustered into four sliding cables with some pulleys being installed at the connection points. To control these sliding cables, as well as the whole structure, more easily, the four lower nodes 1-4 are extended to four external nodes 5-8. Consequently, an actuation of these external nodes is completely equivalent to an actuation of the sliding cables (and the structure). As indicated in Fig. 16 (a), the tower has ten layers with an overall height of 5 m and a width/length of 1 m, at the initial fully unfolded state; each single sliding cable contains 10 pulleys (sliding nodes) and 11 segments; the additional extend lengths of sliding cables are 1 m. Detailed material and dimensional parameters are listed in **Table 1**.

**Table 1** Material and dimensional parameters of the structure

| Parameter | Value |
|---|---|
| Section radius of all cables | 1 mm |
| Young’s modulus of all cables | 73 Gpa |
| Diameters of struts | 3 cm |
| Length of struts | 122.47 cm |
| Young’s modulus of struts | 110 GPa |
| Density of cables | 2450 kg/m$^3$ |
| Density of struts | 4500 kg/m$^3$ |
| friction coefficient | $\mu$ |

At the initial state, no external force or prestress force is applied to the structure, and all the components are in their *non-loaded* lengths. The bottom four nodes 1-4 are all fixed. The middle twenty short classical cables are replaced by light spring elements with a unified axial stiffness of 2e4 N/m, which is specially introduced to fold/unfold such structure with less energy consumption, since these elements are exactly the main deformed components in the actuation process. A cable-based actuation strategy is employed to fold such tower, which is performed in such a way that the four sliding cables are simultaneously stretched by driving nodes 5-8 with a constant and uniform motion speed $V$. These nodes will stop and immediately become locked once a given actuation value is achieved. Actuation value is defined exactly as the motion distance of these nodes. For the deployment of such structure, a reverse process is employed: the sliding cables are loosened progressively and the tower will deploy to the unfolded state, due to the energy stored in spring elements. In this example, only the folding process is considered. In the following analyses, all the struts and the classical cables are modeled by truss elements based on a position description strategy, as detailed in [12], and the sliding cables are analyzed using the approach presented in this work. The actuation behaviors are considered as external constraints and handled in the governing equation by using Lagrange multiplier method. To purely investigate the friction effects induced by the sliding cables, possible collisions between struts are not taken into account; the system thus evolves as if the struts can make a contact of their surfaces or even penetrate each other if the spacing of them is too small. Struts collision analyses will be specifically addressed elsewhere.

4.3.1 Quasi-static folding analysis

We first investigate the influences of friction coefficient on the structural folding performance under the quasi-static actuation strategy, with a slightly large actuation value 1.8 m. The quasi-static actuation strategy assumes that the moving velocities of nodes 5-8 are extremely slow such that inertia effects can be neglected. As a result, a quasi-static analysis is exactly equivalent to a conventional static analysis. Four cases of friction coefficient are investigated: $\mu = 0$, $\mu = 0.05$, $\mu = 0.1$, $\mu = 0.2$. All the analyses are conducted by using Newton-Raphson method, with a convergence error $\xi = 1e\text{-}7$. Fig. 17 gives evolutions of the structural overall height including heights of some internal layers. It is clear that, for all the cases, the height curves decrease monotonously, indicating that the tower is being folded along with the increasing actuation value. On the other hand, the friction coefficient has a significant influence on the structural folding performance. For the frictionless case ($\mu = 0$), the height curves of all the layers decrease smoothly with an equal proportion of the layer index. This is reasonable, because tensions in all segments of the sliding cables are exactly the same, which leads to a uniform contraction of every layer of the structure. While for these frictional cases, results are of distinct diversity. The first (bottom) layer will be first compressed to a zero height, and then the slightly upper layers follow subsequently. In these cases, strut collisions are destined to occur. Generally, with the same actuation value, the larger the friction coefficient, the lower the structural overall height. Some snapshots of configurations corresponding to the actuation value of 1.4 m are displayed in Fig. 18, which clearly demonstrates the non-uniform contraction of each layer for these frictional cases. To gain an insight, Fig. 19 gives the accumulated sliding lengths of some representative pulleys. It can be found that a larger friction coefficient leads to a less sliding length for a particular pulley, which agrees with the well-known fact that frictions tend to prevent sliding motion. Besides, for the first pulley, its accumulated sliding length is almost identical to the actuation value, which is definitely reasonable because the deformation of the sliding cable is small and almost all the actuation value is able to transform into the sliding length. Fig. 20 gives tensions in some segments for the three frictional cases. Apparently, the friction has a severe influence on the internal force transmission: it aggravates tension differences of different segments. For the friction coefficient $\mu$ = 0.2, the maximum tension (in the first segment) is about four times the minimum tension (in the last segment). This may explain why in these frictional cases, the bottom layers of the structure will evolve to a height of zero while the upper layers remain stretched. The above results may lead us to conclude that frictions are detrimental to the folding/unfolding process of such kind of structure, and the friction coefficient should be reduced as much as possible.

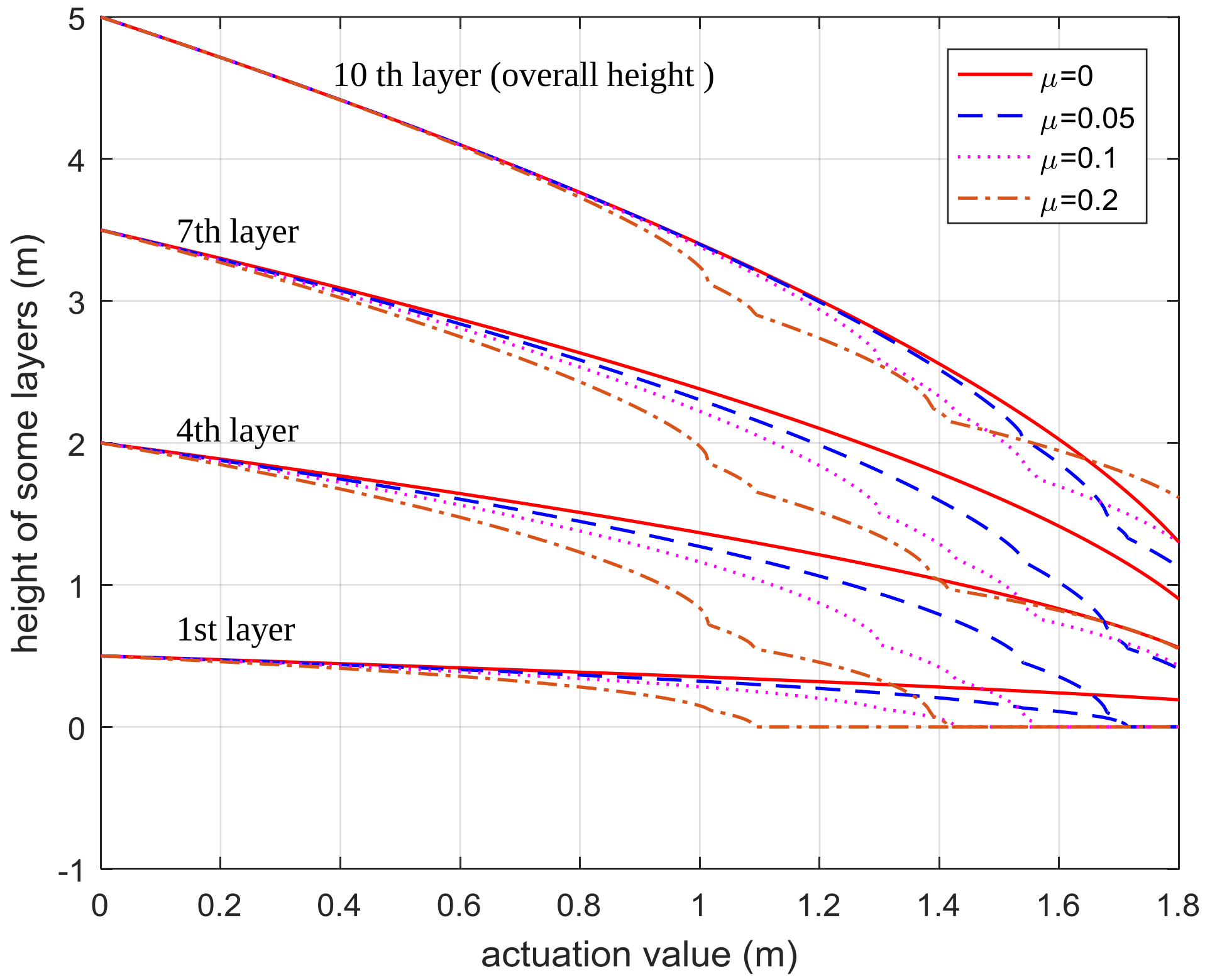

Fig. 17 Evolutions of the structural overall height including heights of some internal layers.

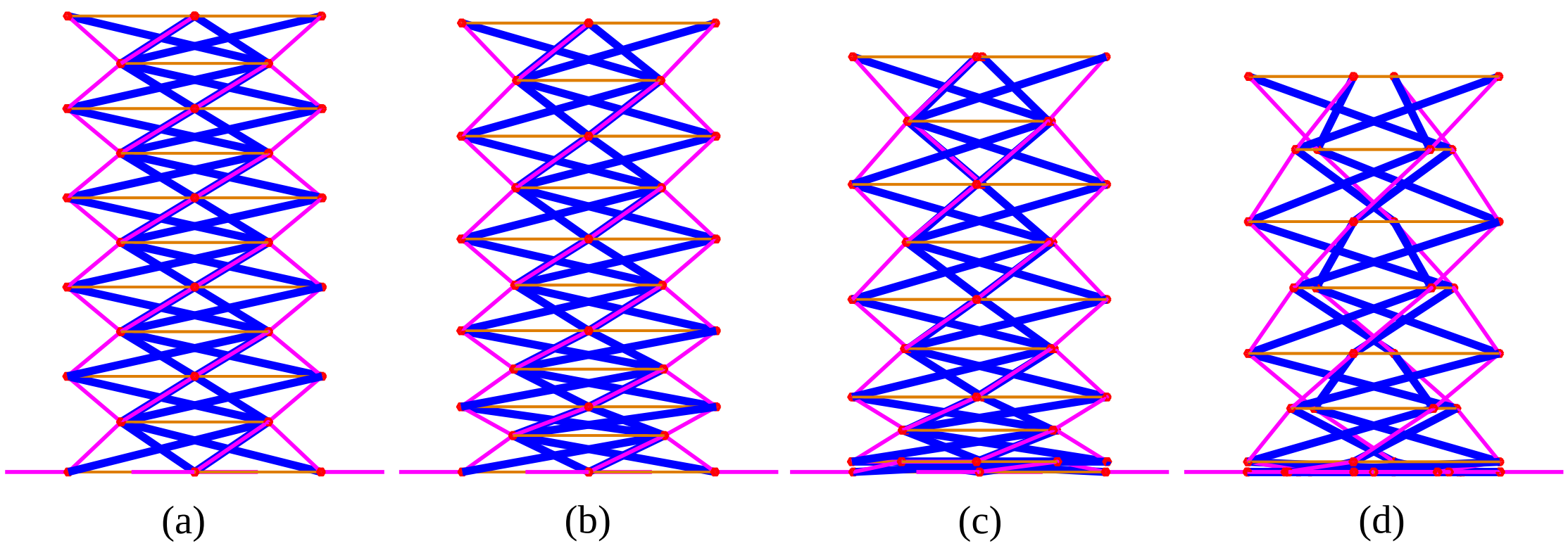

Fig. 18 Snapshots of configurations for different cases of friction coefficient (all correspond to the actuation value of 1.4 m): (a) $\mu$ = 0, (b) $\mu$ = 0.05, (c) $\mu$ = 0.1, (d) $\mu$ = 0.2.

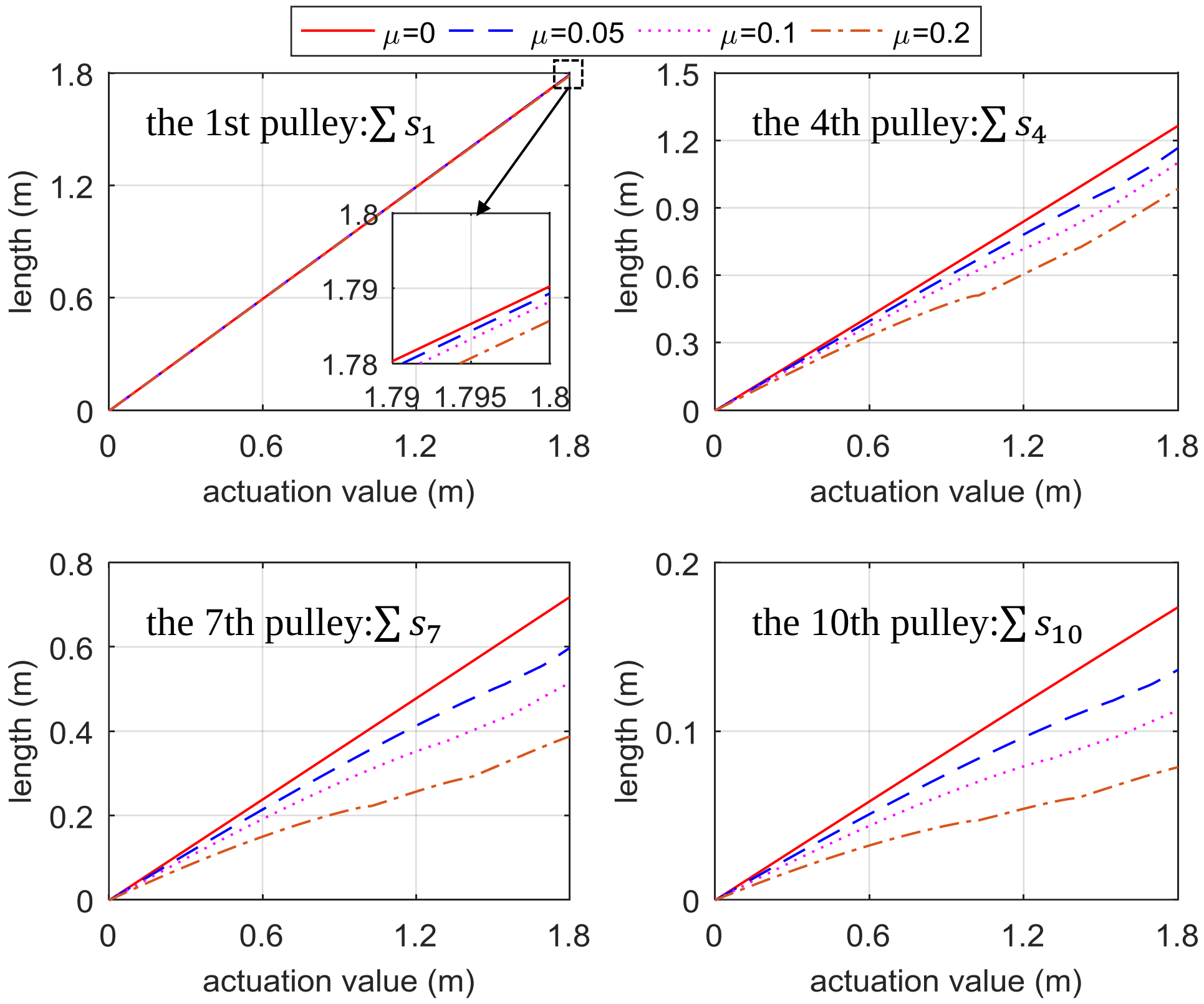

Fig. 19 Accumulated sliding lengths of the 1st, 4th, 7th and 10th pulley

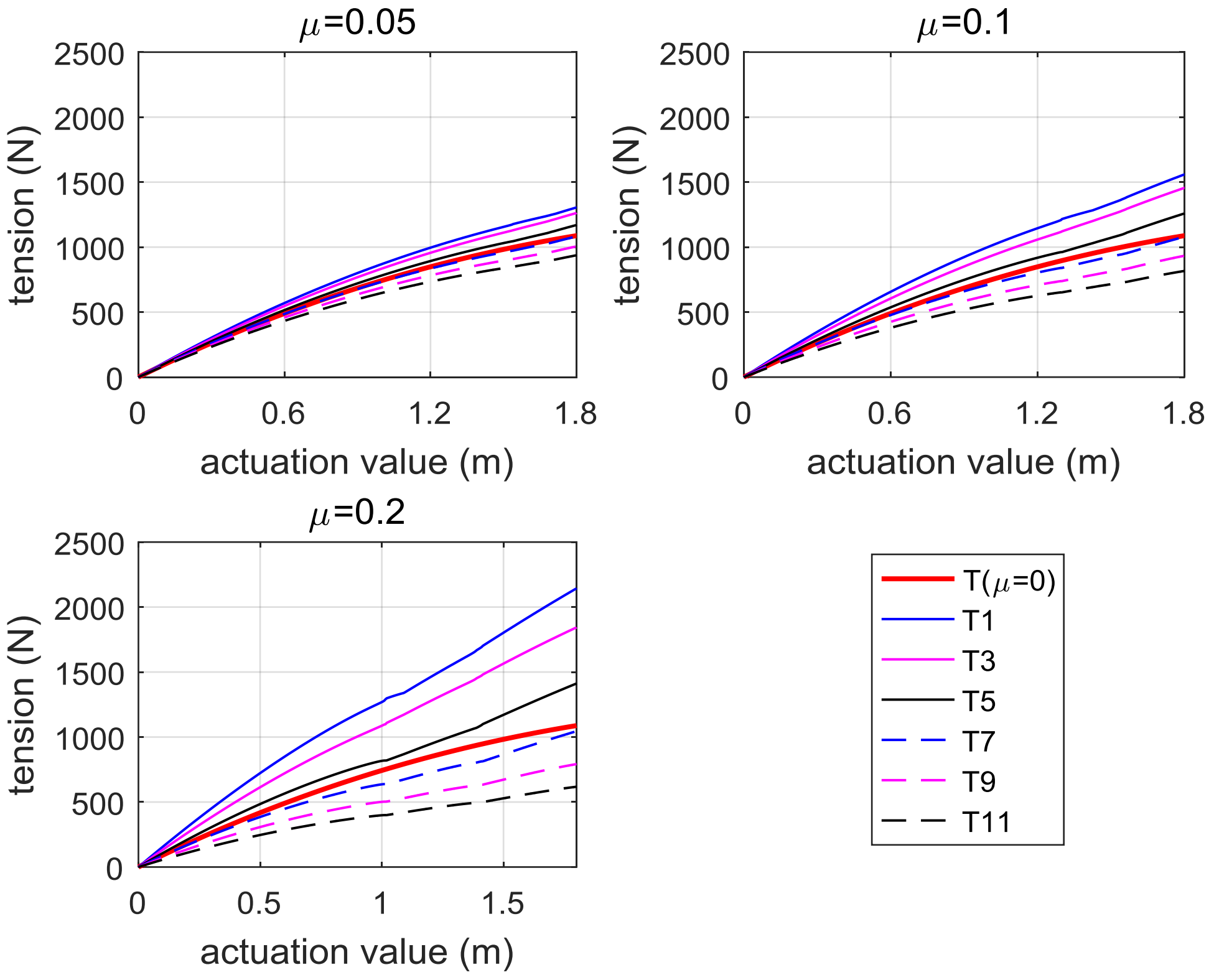

Fig. 20 Tensions in some segments for the three frictional cases, with comparisons to results of the frictionless case.

4.3.2 Dynamic folding analysis

The previous analyses are all conducted by assuming that the moving velocities of nodes 5-8 are extremely slow such that inertia effects can be neglected. However, in real applications, it is usually obscure that to what extent the moving velocities can be regarded as "extremely slow" such that a static analysis procedure can be used. To explore this question, as well as to further demonstrate the ability of the proposed approach in the more complicated dynamic case, we also apply dynamic analyses by considering that nodes 5-8 are all moved with a certain speed $V$ until the designed terminated actuation value has been performed. To avoid excessive non-uniform contraction of the tower, we use a fixed small friction coefficient $\mu$ = 0.05 and set the terminated actuation value as 1.4 m (within this value, the tower can be folded to about half of its original height, as indicated in Fig. 17). Four cases of actuation speed are considered: $V$ = 1, 0.5, 0.1, 0.05 m/s; the corresponding folding times are 1.4, 2.8, 14 and 28 s, respectively. For all these cases, dynamic analyses are carried out using the Newmark algorithm with parameters $\alpha$ = 0.3, $\delta$ = 0.5 (symbolic definitions are coincide with [34]), with a time step size $\Delta t$=0.001 s and a total analysis time 60 s. Time intervals beyond the folding times are used to observe structural vibration characteristics.

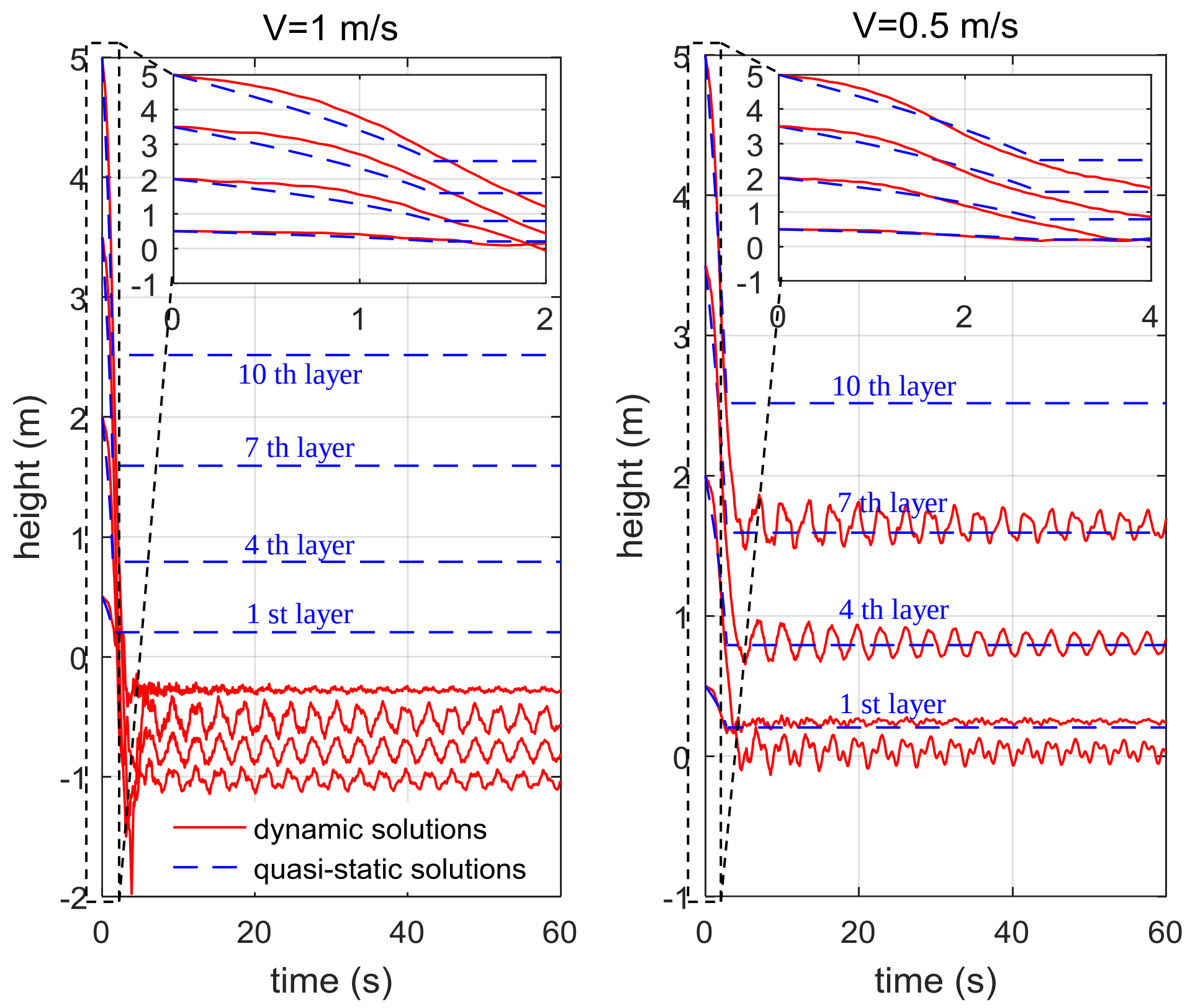

Fig. 21 Time history curves of heights of some layers for $V$ = 1 m/s and 0.5 m/s.

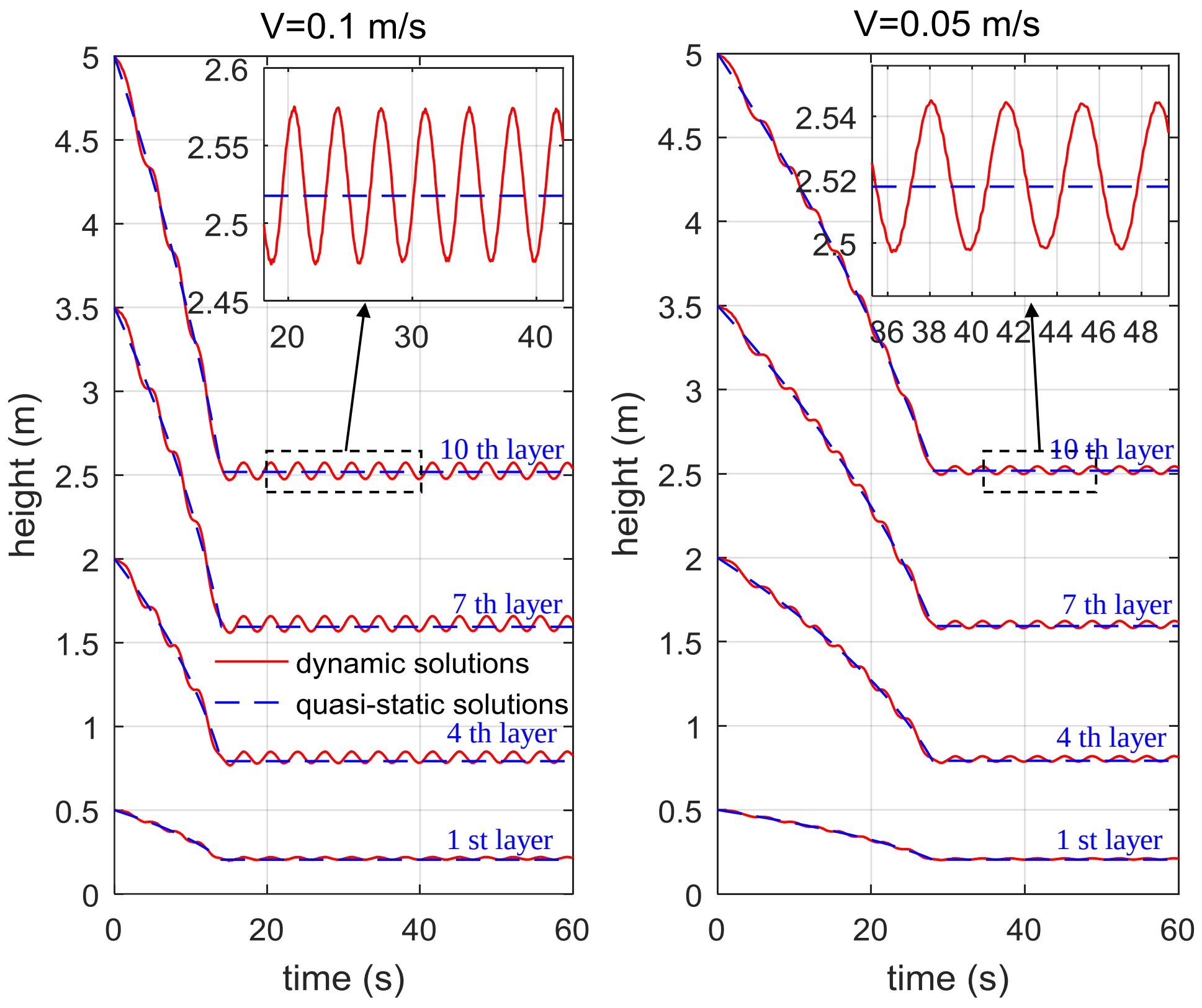

Fig. 22 Time history curves of heights of some layers for $V$ = 0.1 m/s and 0.05 m/s

General behaviors of the dynamic folding actuation are first investigated. Fig. 21 and Fig. 22 give time history curves of heights of some layers for the four actuation speed cases. The corresponding quasi-static results are also included for comparisons. It is clear that for the first two speed cases, the height curves are deviated from the quasi-static results, and some layers even achieve to a zero or negative height, implying that some struts have penetrated each other. These results demonstrate that the dynamic effects are significant for these levels of actuation speed. While for the last two speed cases, the height curves are very close to their corresponding quasi-static solutions. The structure will be folded to the designated height, and then vibrate slightly around the final folded state when actuation is terminated. These curves further imply that $V = 0.1$ m/s may be an acceptable choice of actuation speed, on one hand, to pursue a fast folding process, on the other hand, to avoid inducing any complex behavior involving struts collision.

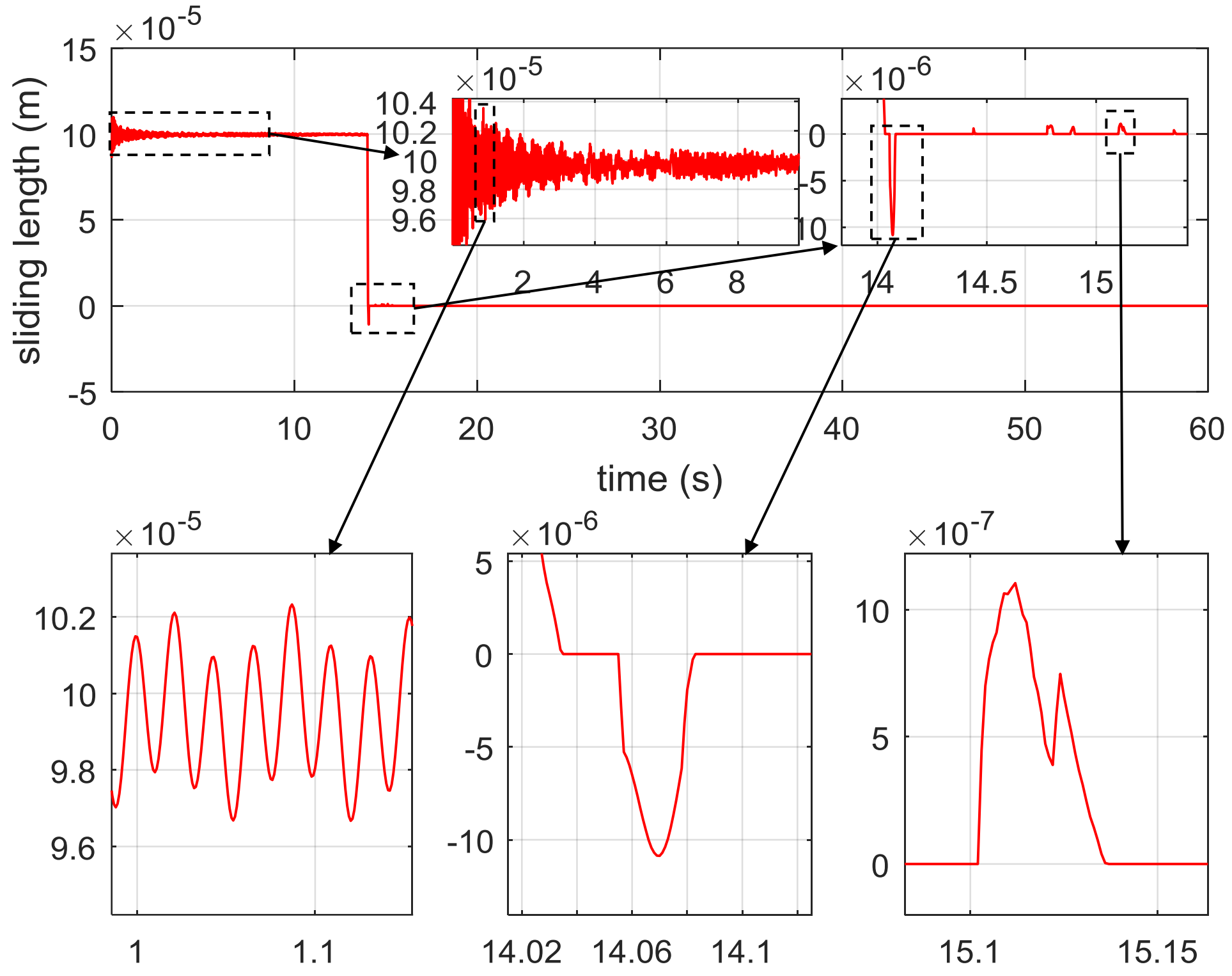

Fig. 23 Time history curve of sliding length for the first pulley including its enlarged drawings.

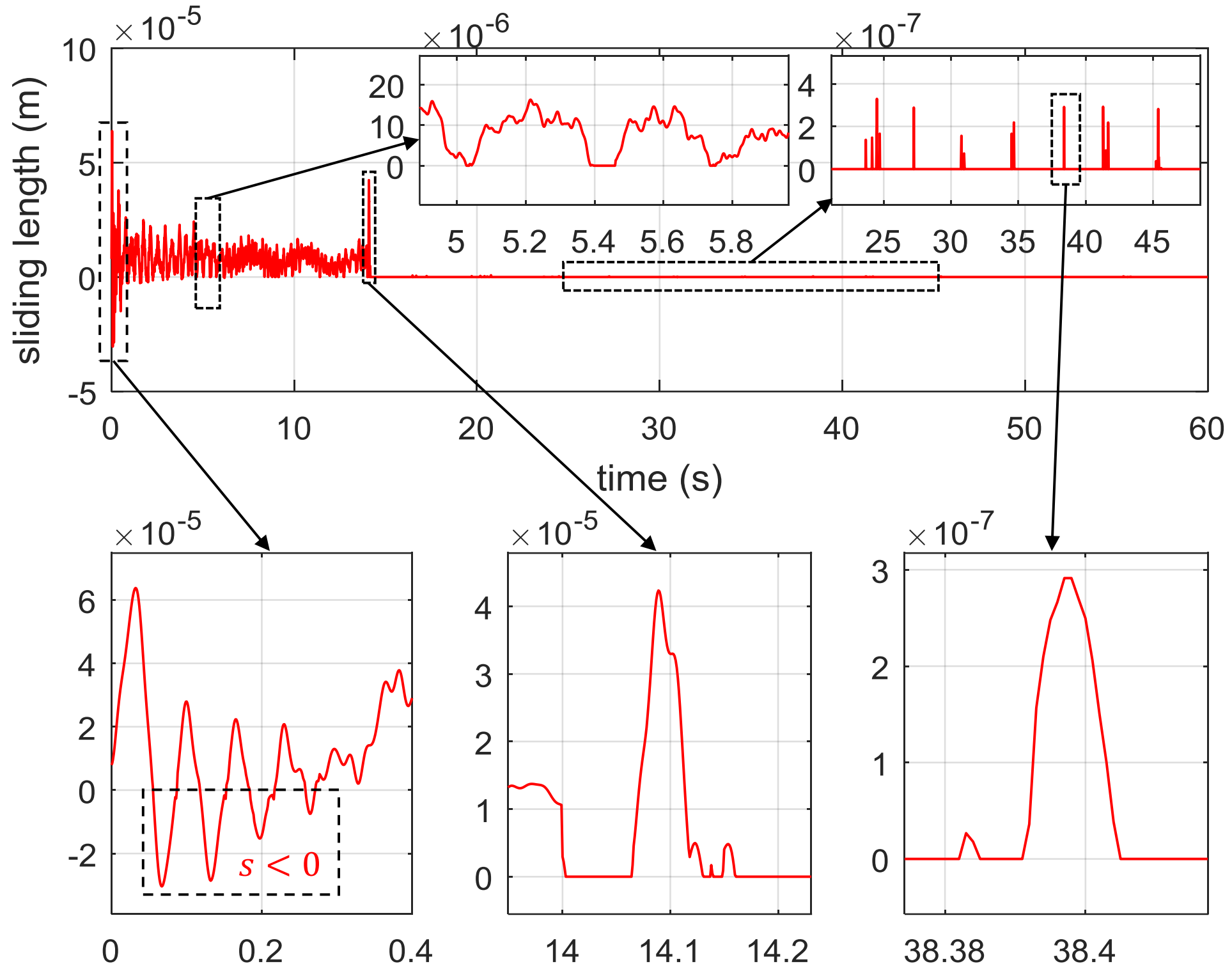

Fig. 24 Time history curve of sliding length for the last (10th) pulley including its enlarged drawings.

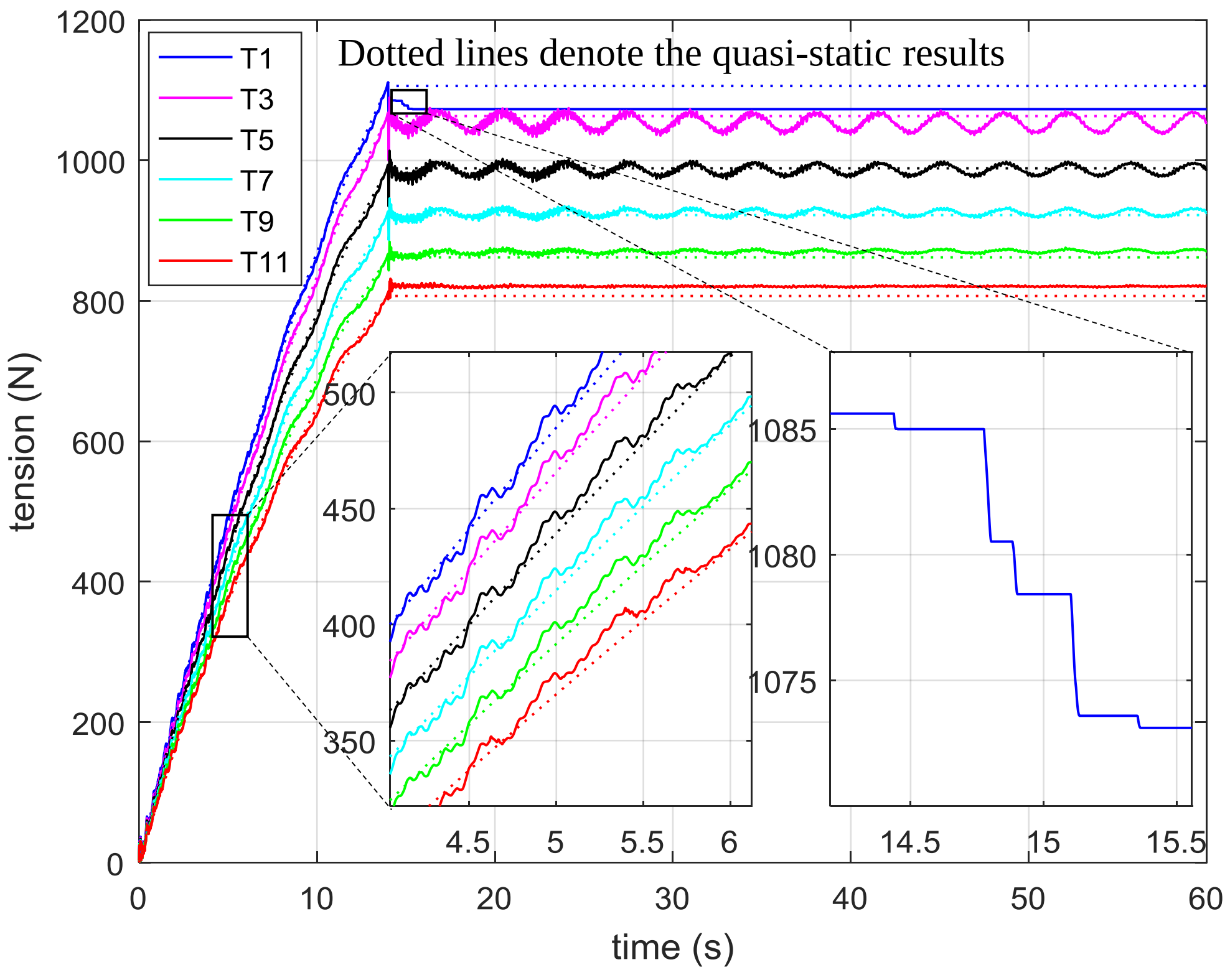

Fig. 25 Time history curves of tensions in some representative segments including their corresponding results of the quasi-static case.

Next, we are interested in how these pulleys move in the dynamic actuation process. Using results of the actuation speed $V = 0.1$ m/s, Fig. 23 and Fig. 24, respectively, give time history curves of the (one-step) sliding length for the first and the last (10th) pulley, including their enlarged drawings. It is clear to see that the motions of the pulleys can be divided into two distinct phases: $0 < t < 14$ s, $14 < t < 60$ s. In the first phase, the actuation is being performed and the structure is being folded gradually. Almost all the time, the pulleys slide to the end node (as $s > 0$), but there are still some short time intervals in which some pulleys experience a reverse sliding ($s < 0$). In the second phase, the actuation is terminated and the previously moving nodes are all locked, and the structure is in a state of free vibration. Motion states of the pulleys are much more diverse: most of the time, the pulleys cling to the cables (sticking state, $s = 0$); sometimes they may slide to the end node ($s > 0$), but rarely slide to the first node ($s < 0$). Fig. 25 gives time history curves of tensions in some representative segments, as well as results of the quasi-static case. In the first actuation phase, since the actuation speed is not very large, and dynamic effects are not significant, the tension curves evolve generally along the quasi-static results. In the second phase, the tensions vibrate. Besides, it seems that the tension differences for different segments are slightly weakened compared to the quasi-static results, perhaps due to the intermittent pulley sliding effects induced by the vibration. It is particularly interesting to note from the enlarged drawing of Fig. 25 that tension of the first segment in the vibration phase has some step-like descents, which is absolutely reasonable. For such segment, once the actuation is terminated, its two ends are all fixed; the subsequent tension is governed purely by the accumulated sliding length of the first pulley. Each descent of the tension curve represents an occurrence of sliding motion of such pulley (towards the end node).

## 5. Conclusions

Sliding cable system with frictions is encountered in many engineering applications. Such system is typically characterized by existences of diversified motion states of different sliding nodes (pulleys), which leads to significant difficulties to analyze. This paper presents a very simply linear complementarity approach for analysis of sliding cable with frictions. Within this approach, the challenging and significant issue of determinations of motion states of sliding nodes, as well as calculations of their sliding lengths, can be handled in a standard linear complementarity formulation. The merit of this approach is twofold:

(1) From a numerical point of view, the proposed approach opens a possibility of applying any available efficient LCP solver in the analysis, eliminating the need for traditional cumbersome predictor-corrector operation. It also allows us to write the program in a very compact two-step form: the first one is for the LCP solving and the second one is to obtain the final nodal force vectors using solutions of the LCP.

(2) From a theoretical point of view, the proposed approach opens a way to study the theory problem such as the existence and uniqueness of the sliding lengths by using complementarity theories. The uniqueness of the solution has been proved in this work, while the existence issue remains open. Further inquiries to this issue may provide us a profound insight into sliding behaviors of the sliding nodes.

Multi examples involving both static and dynamic analyses are presented to demonstrate the effectiveness of the proposed approach. The results, from multiple perspectives, highlight the complicated mechanical behaviors, in particular the variegated motion states of the sliding nodes, involved in sliding cables when taking into account frictions. Despite the complexity, the problem can still be handled very well using the proposed approach. The proposed approach, we think, is promising to be a popular method in dealing with multi-node sliding cable with frictions, given its simplicity and effectiveness.

## Acknowledgments

The authors are grateful for the financial support of the National Key Research and Development Plan (2016YFB0200702); the National Natural Science Foundation of China (11772074, 11472069, 11761131005, and 91515103); the National 111 Project of China (B14013). The authors also would like to thank Prof. Nizar Bel Hadj Ali from University of Carthage, Tunisia, for the valuable discussions on sliding cable modeling with frictions.

## References

[1] F. Ju, Y.S. Choo, Super element approach to cable passing through multiple pulleys, Int J Solids Struct, 42 (2005) 3533-3547.

[2] M. Aufaure, A finite element of cable passing through a pulley, Comput Struct, 46 (1993) 807-812.

[3] M. Aufaure, A three-node cable element ensuring the continuity of the horizontal tension; a clamp–cable element, Comput Struct, 74 (2000) 243-251.

[4] B. Zhou, M.L. Accorsi, J.W. Leonard, Finite element formulation for modeling sliding cable elements, Comput Struct, 82 (2004) 271-280.

[5] K.H. Lee, Y.S. Choo, F. Ju, Finite element modelling of frictional slip in heavy lift sling systems, Comput Struct, 81 (2003) 2673-2690.

[6] J.B. Coulibaly, M.A. Chanut, S. Lambert, F. Nicot, Sliding cable modeling: An attempt at a unified formulation, Int J Solids Struct, 130-131 (2018) 1-10.

[7] Z.H. Chen, Y.J. Wu, Y. Yin, C. Shan, Formulation and application of multi-node sliding cable element for the analysis of Suspen-Dome structures, Finite Elem Anal Des, 46 (2010) 743-750.

[8] K.W. Moored, H. Bart-Smith, Investigation of clustered actuation in tensegrity structures, Int J Solids Struct, 46 (2009) 3272-3281.

[9] N. Bel Hadj Ali, L. Rhode-Barbarigos, I.F.C. Smith, Analysis of clustered tensegrity structures using a modified dynamic relaxation algorithm, Int J Solids Struct, 48 (2011) 637-647.

[10] L. Zhang, M.K. Lu, H.W. Zhang, B. Yan, Geometrically nonlinear elasto-plastic analysis of clustered tensegrity based on the co-rotational approach, Int J Mech Sci, 93 (2015) 154-165.

[11] L. Zhang, Q. Gao, Y. Liu, H. Zhang, An efficient finite element formulation for nonlinear analysis of clustered tensegrity, Eng Computation, 33 (2016) 252-273.

[12] Z. Kan, H. Peng, B. Chen, W. Zhong, Nonlinear dynamic and deployment analysis of clustered tensegrity structures using a positional formulation FEM, Compos Struct, 187 (2018) 241-258.

[13] Z. Kan, H. Peng, B. Chen, W. Zhong, A sliding cable element of multibody dynamics with application to nonlinear dynamic deployment analysis of clustered tensegrity, Int J Solids Struct, 130-131 (2018) 61-79.

[14] H. Liu, Q. Han, Z. Chen, X. Wang, R.-Z. Yan, B. Zhao, Precision control method for pre-stressing construction of suspen-dome structures, Advanced Steel Construction 10 (2014) 404-425.
[15] N. Veuve, S.D. Safaei, I.F.C. Smith, Active control for mid-span connection of a deployable tensegrity footbridge, Eng Struct, 112 (2016) 245-255.
[16] N. Veuve, S.D. Safaei, I.F.C. Smith, Deployment of a Tensegrity Footbridge, Journal of Structural Engineering, 141 (2015) 04015021.
[17] Y.S. Choo, F. Ju, K.H. Lee, Static Sling Tensions in Heavy Lifts with Doubled Sling Arrangement, International Journal of Offshore and Polar Engineering, 7 (1997) 313-322.
[18] K. Hincz, Nonlinear Analysis of Cable Net Structures Suspended from Arches with Block and Tackle Suspension System, Taking into Account the Friction of the Pulleys, International Journal of Space Structures, 24 (2009) 143-152.
[19] T. Erhart, Pulley mechanism for muscle or tendon movements along bones and around joints. In: LS-DYNA Forum. DYNAmore,, (2012).
[20] N. Bel Hadj Ali, A.C. Sychterz, I.F.C. Smith, A dynamic-relaxation formulation for analysis of cable structures with sliding-induced friction, Int J Solids Struct, 126-127 (2017) 240-251.
[21] R.W. Cottle, J.-S. Pang, R.E. Stone, The Linear Complementarity Problem, Boston : Academic Press, 1992.
[22] H.W. Zhang, S.Y. He, X.S. Li, Two aggregate-function-based algorithms for analysis of 3D frictional contact by linear complementarity problem formulation, Comput Method Appl M, 194 (2005) 5139-5158.
[23] H.W. Zhang, J.Y. Li, S.H. Pan, New second-order cone linear complementarity formulation and semi-smooth Newton algorithm for finite element analysis of 3D frictional contact problem, Comput Method Appl M, 200 (2011) 77-88.
[24] Y. Kanno, M. Ohsaki, A non-interior implicit smoothing approach to complementarity problems for frictionless contacts, Comput Method Appl M, 200 (2011) 1176-1185.
[25] G. Bolzon, G. Maier, F. Tin-Loi, On multiplicity of solutions in quasi-brittle fracture computations, Comput Mech, 19 (1997) 511-516.
[26] A. Bassi, N. Aravas, F. Genna, A Linear Complementarity formulation of rate-independent finite-strain elastoplasticity. Part I: Algorithm for numerical integration, European Journal of Mechanics - A/Solids, 35 (2012) 119-127.
[27] C. Hager, B.I. Wohlmuth, Nonlinear complementarity functions for plasticity problems with frictional contact, Comput Method Appl M, 198 (2009) 3411-3427.
[28] N.D. Oliveto, M.V. Sivaselvan, Dynamic analysis of tensegrity structures using a complementarity framework, Comput Struct, 89 (2011) 2471-2483.
[29] L. Zhang, Q. Gao, H.W. Zhang, An efficient algorithm for mechanical analysis of bimodular truss and tensegrity structures, Int J Mech Sci, 70 (2013) 57-68.
[30] C.E. Lemke, Bimatrix Equilibrium Points and Mathematical Programming, Manage Sci, 11 (1965) 681-689.
[31] F.B. Beer, E.R. Johnston, Vector Mechanics for Engineers: Statics and Dynamics, McGraw-Hill, New York, 1996.
[32] A.R. Conn, N.I.M. Gould, P.L. Toint, Trust Region Methods, 2000.
[33] M. Barnes, Form finding and analysis of tension structures by dynamic relaxation, International Journal of Space Structures, 14 (1999) 89-104.
[34] K.-J. Bathe, Finite Element Procedures, Prentice Hall, Upper Saddle River, New Jersey 1996.

[35] R.E. Skelton, M.C. De Oliveira, Tensegrity Systems, Springer, 2009.